\documentclass[12pt]{article}
\usepackage{amssymb}
\usepackage{amsmath}
\usepackage{amsthm}
\usepackage{amsbsy}
\usepackage{amsfonts}
\usepackage[cp1251]{inputenc}
\usepackage[english]{babel}

\usepackage{graphicx}

\newtheorem{thm}{Theorem} [section]

\newtheorem{prop}[thm]{Proposition}

\newtheorem{rem}[thm]{Remark}

\numberwithin{equation}{section}

\newcommand{\cK}{\mathcal {K}}

\newcommand{\cL}{\mathcal {L}}
\newcommand{\cH}{\mathcal {H}}

\newcommand{\cT}{\mathcal {T}}
\newcommand{\cF}{\mathcal {F}}

\newcommand{\cR}{\mathcal {R}}

\newcommand{\bR}{\mathbb {R}}
\newcommand{\bZ}{\mathbb {Z}}
\newcommand{\bC}{\mathbb {C}}
\newcommand{\bN}{\mathbb {N}}

\newcommand{\bU}{\mathbb {U}}
\newcommand{\bT}{\mathbb{T}}

\newcommand{\be}{-e}
\newcommand{\trans}{\cT}
\newcommand{\sgn}{{\rm sgn }}

\newcommand{\rRe}{{\rm Re\,}}
\newcommand{\rIm}{{\rm Im\,}}

\newcommand{\coker}{{\rm coker\,}}
\newcommand{\rdiv}{\mathrm{div}}

\newcommand{\Res}{{\rm Res}}
\newcommand{\bW}{\mathbb{W}}
\newcommand{\rran}{\mathrm{ran}\,}

\newcommand{\exnum}{d}

\begin{document}
	\begin{center}
		\begin{Large}
			{\bf Radiation principle for discrete acoustic equation on graphs with several cylindrical outlets to infinity}
		\end{Large}
		
		\vspace{4ex}
		\begin{large}
			A.~S. Poretskii, D.~S. Smorchkov 
		\end{large}
		\footnote{The research was supported by the Russian Science Foundation grant No. 25-21-00533} 
	\end{center}
	\begin{abstract}
		A discrete analogue of the acoustic equation with a spectral parameter is studied on graphs with several cylindrical outlets to infinity. At infinity in each cylindrical outlet the coefficients of this equation converge exponentially fast to periodic functions. It is assumed that the spectral parameter is fixed and distinct from thresholds and flat bands. We construct a special basis of the continuous spectrum eigenfunctions (bounded solutions of the homogeneous equation that do not belong to $\ell_{2}$), introduce the scattering matrix, and establish the radiation principle (a well-posed statement of the problem with radiation conditions at infinity).
	\end{abstract}
	\section{Introduction}
	\begin{figure}[h!]\centering
		\includegraphics[width=0.28\textwidth]{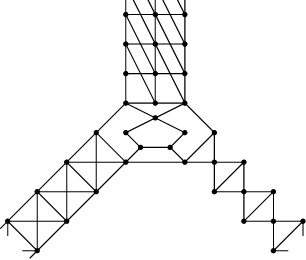}
		\caption{Example of a branched waveguide}\label{fWG}
	\end{figure}
	
	This article is devoted to studying an acoustic equation on discrete waveguides. A simple example of a discrete waveguide is a discrete cylinder, i.e., a graph that is periodic under a shift by a given vector and has a finite periodicity cell. We study branched waveguides that are graphs with several cylindrical outlets to infinity (see Fig. \ref{fWG}). 
	On such a graph we consider the discrete acoustic equation
	\begin{align}\label{eq_A_explicit_Intro}
		\sum\limits_{e\in E(x)}a(e)\big(u(x)-u(t(e))\big)+q(x)u(x)-\mu u(x)=f(x), \quad x\in V.
	\end{align}
	Here the unknown $u$ and the right-hand side $f$ are functions on the set $V$ of vertices of the graph, the summation is over all edges of the graph incident to the vertex $x$, and $t(e)$ stands for the other endpoint of the edge $e$. The coefficients $a(e)$ and $q(x)$ are functions defined on the sets of edges and vertices of the graph, respectively. It is assumed that at infinity in each cylindrical outlet $a(e)$ and $q(x)$ converge exponentially fast to functions periodic under a shift by the period of the corresponding cylinder (we call this condition stabilization of the coefficients at infinity). 
	Equation \eqref{eq_A_explicit_Intro} is the discrete analogue of the differential equation $-\mathrm{div}a\nabla u+qu=f$ and includes the discrete Schr\"odinger equation as the special case $a\equiv 1$. Equation \eqref{eq_A_explicit_Intro} may be supplemented by a Dirichlet condition 
	\begin{align}\label{eq_Dir_Intro}
		u(x)=g(x), \quad x \in \partial V,
	\end{align}
	where $\partial V \subset V$ is a set of vertices that plays the role of the waveguide boundary. In this case equation \eqref{eq_A_explicit_Intro} is considered for $x \in \mathring{V}:=V\setminus\partial V$. The main goals of the work are to establish the radiation principle (a well-posed statement of problem \eqref{eq_A_explicit_Intro}, \eqref{eq_Dir_Intro} with intrinsic radiation conditions at infinity) and to introduce the scattering matrix.  The article assumes that the spectral parameter $\mu \in \bR$ is fixed, belongs to the continuous spectrum, and is distinct from thresholds and flat bands (definitions are given in Section \ref{ss_Fund_spec}), but it may be an eigenvalue of finite multiplicity.
	
	Branched discrete waveguides can be considered as a mathematical model of quasi-one-dimensional crystals with defects. Another possible application is related to numerical modelling of similar ``continuous'' waveguides described by the acoustic equation on domains in $\bR^d$ with several cylindrical outlets to infinity. Discrete problems are better suited for computing, and their solutions allow one to construct approximations of solutions to the corresponding boundary-value problems for differential equations. Finally, the theory of continuous waveguides is well developed, and constructing a parallel theory of discrete waveguides is a relevant task. At the same time, new questions arise, related, for example, to the discrete interpretation of differential operators or to the peculiarities of graph topology. In addition, discrete problems provide nontrivial examples of explicitly solvable problems in waveguide theory. 
	
	Let us give an overview of mathematical research works in this area. Most of them are devoted to the discrete Schr\"odinger equation on graphs with various structure. In the simplest case, the waveguide is the set of vertices $\bZ$ connected successively by edges. This case has been studied in detail, see \cite{EgorKopTes}. However, for arbitrary periodic graphs there is no single completed theory. One direction of research is concerned with studying the properties of the spectrum for the periodic discrete Schr\"odinger operator on such graphs. Kotani, Sunada, and coauthors developed a geometric approach to spectral analysis of abelian coverings for finite graphs, see \cite{Su13} and references therein. Higuchi and Shirai \cite{HS} applied the approach to discrete magnetic Schr\"odinger operator, relating the operator on a periodic graph to a family of operators on a finite fundamental graph (quotient graph). 
	Korotyaev and Saburova extended this theory to discrete Schr\"odinger operators with a periodic potential and obtained more detailed information about the band structure of the spectrum and its dependence on the geometry of the fundamental graph, see e.g. \cite{KorSab1}.
	We also mention the work \cite{SloKor}, where the Schr\"odinger operator with a potential vanishing at infinity was studied and an asymptotics of the counting function for the discrete spectrum was established.
	
	Papers of the second direction are devoted to the scattering theory, where the main task is to describe the evolution of states corresponding to the continuous spectrum of the system. In \cite{AndIsoMar}, the periodic Schr\"odinger operator on a periodic graph is considered, and local perturbations (on a finite part of the graph) of the coefficients and the graph geometry are allowed. The continuous spectrum eigenfunctions are constructed, the scattering matrix is introduced, and the existence and completeness of the wave operators are proved. In \cite{ParaRich}, the acoustic equation is considered under much milder restrictions on the perturbations of the coefficients (admitting power-law convergence rate), but the graph remains periodic. The existence and completeness of the wave operators are proved, but the scattering matrix is not introduced there. Similar questions were discussed in \cite{Para} for the Gauss--Bonnet operator. The term ``Schr\"odinger equation on graphs'' can also refer to other types of problems, in particular, to ``metric graphs''; in this connection see \cite{BerKuch}.
	
	Our approach is based on the waveguide theory for elliptic boundary-value problems in domains with several cylindrical outlets to infinity, which is well developed. The foundation of this theory was laid in the monograph \cite{NazPlam}, where, in particular, the radiation principle was described and justified and the scattering matrix was introduced.The main research scheme from this theory remains applicable to problems for difference equations in discrete waveguides.
		
	Let us describe the structure of the article and formulate the main results. The first part is the Introduction. The second part gives necessary definitions from general graph theory, introduces the main function classes, and describes the discrete analogues of the gradient and divergence operators, the acoustic operator, and various boundary conditions. Then a discrete cylinder is constructed, its fundamental graph is defined, and the notion of the edge index is introduced. On the cylinder, the Fourier transform is introduced.
	
	Part 3 is devoted to the study of system \eqref{eq_A_explicit_Intro}, \eqref{eq_Dir_Intro} with periodic coefficients in a discrete cylinder. By means of the Fourier transform, this problem is reduced to studying the spectrum of a holomorphic operator-valued function corresponding to a problem on the fundamental graph. Based on the information obtained, an asymptotics of the solutions to the problem in the cylinder is described, sets of incoming and outgoing waves $\{u_{j}^{+}\}$ and $\{u_{l}^{-}\}$ are introduced.
	
	Finally, in Part 4, problem \eqref{eq_A_explicit_Intro}, \eqref{eq_Dir_Intro} with coefficients stabilizing at infinity is considered on a graph with several cylindrical outlets. The operator of this problem is studied in function classes with exponential weight, and its Fredholm property is proved (Theorem \ref{th:3.1.1}). 
	The asymptotics of an arbitrary solution to problem \eqref{eq_A_explicit_Intro}, \eqref{eq_Dir_Intro} at infinity is expressed in terms of the incoming and outgoing waves. The main results are derived from these theorems. The first one is related to choosing a basis of the continuous spectrum eigenfunctions. In Theorem \ref{th:3.3.6}, the existence of the continuous spectrum eigenfunctions $Y_{j}^{+}$ with the asymptotics
	\begin{align}\label{eq:intro1}
		\begin{gathered}
			Y_{j}^{+}=u_{j}^{+}+\sum\limits_{l=1}^{\Upsilon}u_{l}^{-}S_{lj}+o(1),\quad j=1,\ldots,\Upsilon,
		\end{gathered}
	\end{align}
	is established. Here $\Upsilon$ is the number of incoming (outgoing) waves, and it is called the multiplicity of the continuous spectrum. The matrix $S$ with entries $S_{lj}$ is unitary and is called the scattering matrix. The second main result of the paper is the radiation principle. It provides a well-posed statement of the problem with radiation conditions at infinity (Theorem \ref{th:3.3.7}). These conditions specify the type of asymptotics of the solution at infinity:
	\begin{align}\label{eq:intro2}
		u=c_{1}u_{1}^{-}+\ldots+c_{\Upsilon}u_{\Upsilon}^{-}+o(1).
	\end{align}
	The leading term of the asymptotics consists only of outgoing waves; therefore the radiation conditions \eqref{eq:intro2} are called intrinsic. The problem \eqref{eq_A_explicit_Intro}, \eqref{eq_Dir_Intro} with the radiation conditions \eqref{eq:intro2} is Fredholm of index zero. 
	Assume that the number $\mu$ is not an eigenvalue of problem \eqref{eq_A_explicit_Intro}, \eqref{eq_Dir_Intro}, i.e., the corresponding homogeneous problem has no nontrivial solutions in $\ell_{2}(V)$. Then, for any exponentially decaying right-hand side $\{f,g\}$, the problem \eqref{eq_A_explicit_Intro}, \eqref{eq_Dir_Intro} has a unique solution $u$ satisfying the asymptotics \eqref{eq:intro2}. This solution depends continuously on the right-hand side in suitable function classes. The coefficients $c_{j}$ in the asymptotics \eqref{eq:intro2} can be expressed in terms of the right-hand side and the continuous spectrum eigenfunctions.
	In the case where $\mu$ is an eigenvalue, the problem has a solution if the right-hand side satisfies solvability conditions. The number of such conditions equals the number of linearly independent eigenfunctions corresponding to the value $\mu$.

\section{Preliminaries}
\subsection{Graphs}\label{ssGraph}
Let $V$ be a set of points that we shall call \textit{vertices}. A pair of vertices may be connected by an \textit{edge}. 
We shall call such vertices \textit{adjacent} and say that they are \textit{incident} to this edge or, in other words, are the endpoints of this edge.
An edge is called a \textit{loop} if its endpoints coincide. Edges are called \textit{multiple} if the pairs of their endpoints coincide. 
It will be more convenient for us to use oriented edges, which we shall call \textit{arrows}. To each edge we formally assign two arrows with opposite orientations. We shall denote these arrows by $e$ and $\be$. 
Let $E$  be the set of all possible arrows.
We shall call the pair $G=(V,E)$ a \textit{graph}. From now on we assume that the sets $V$ and $E$ are at most countable. The graph $G=(V, E)$ is called finite if the sets $V$ and $E$ are finite.

To specify the orientation of an edge connecting distinct vertices $x$ and $y$, one has to choose which of these vertices will be the initial vertex $o(e)$ of the corresponding arrow $e$ and which will be its terminal vertex $t(e)$. Suppose $o(e)=x$ and $t(e)=y$; then $o(\be)=y$ and $t(\be)=x$. In this case we shall say that the arrow $e$ \textit{leaves} the vertex $x$ and \textit{enters} the vertex $y$. The arrows $e$ and $\be$ corresponding to a loop incident to the vertex $x$ have $o(e)=t(e)=o(\be)=t(\be)=x$ and differ only formally. 
Let $E(x)$ denote the set of arrows leaving $x$:
\begin{align}\label{eq_E(x)}
	E(x)=\{e\in E: x=o(e)\}.
\end{align}
The \textit{degree} of a vertex $x$ is the number $\deg(x)=|E(x)|$; here and throughout $|F|$ stands for the cardinality of the set $F$. In other words, $\deg(x)$ is the number of edges incident to $x$, while a loop incident to $x$ is counted twice in $\deg(x)$. In what follows, we assume that
\begin{align}\label{eq:1.1.1}
	\deg(G):=\sup\limits_{x\in V}\deg(x)<\infty.
\end{align}

A \textit{morphism} $\varphi$ from a graph $G_1=(V_1,E_1)$ to a graph $G_2=(V_2,E_2)$ is a pair $\varphi=(\varphi_V, \varphi_E)$ of maps $\varphi_V:V_1\to V_2$, $\varphi_E:E_1\to E_2$, satisfying the conditions 
\begin{align}\label{eq_morph}
	\varphi_V(o(e))=o(\varphi_E(e)),\quad \varphi_V(t(e))=t(\varphi_E(e)), \quad -\varphi_E(e)=\varphi_E(-e)
\end{align}
for every arrow $e \in E_1$. A morphism $\varphi$ is called an \textit{isomorphism} if the maps $\varphi_V$ and $\varphi_E$ are bijections; in this case the graphs $G_1$ and $G_2$ are called \textit{isomorphic}.

An ordered sequence of arrows $(e_1,e_2,...,e_{n})$ is called a \textit{walk} if $t(e_i)=o(e_{i+1})$ for all $i=1,\ldots,n-1$; the number $n$ is called the length of the walk.  
A walk is called a \textit{geodesic} if $e_i\neq -e_{i+1}$ for all $i=1, \dots, n-1$; any walk can be shortened to a geodesic by removing all pairs $(e_i, e_{i+1}): e_i= -e_{i+1}$. A graph is called \textit{connected} if any two of its vertices can be joined by a geodesic. A \textit{cycle} is a closed geodesic, i.e., a geodesic such that $t(e_n)=o(e_1)$ and $e_n\neq-e_1$. A \textit{tree} is a connected graph having no cycles; in a tree any pair of vertices can be joined by a unique geodesic. A tree $G_0=(V_0,E_0)$ is called a \textit{spanning tree} of the graph $G=(V,E)$ if $V_0=V$ and $E_0\subset E$.
\subsection{Difference equations and boundary-value problems on a graph}
Let, as before, $G=(V,E)$ be a graph. Denote by $H(V)$ the Hilbert space of functions from $V$ to $\mathbb{C}$ with the inner product
\begin{align}\label{eq:1.2.1}
	(u_{1},u_{2})_{H(V)}=\sum\limits_{x\in V}u_{1}(x)\overline{u_{2}(x)}.
\end{align}
A function $\omega: E \to \bC$ satisfying the equality $\omega(\be)=-\omega(e)$ for every arrow $e \in E$ is called an \textit{odd} function on $E$; here, as before, $e$ and $\be$ are the arrows corresponding to the two possible orientations of one edge. It is sufficient to define the function $\omega$ on a set $E^o\subset E$ containing only one arrow for each edge. Let $H(E)$ stand for the space of odd functions on $E$ with the inner product
\begin{align}\label{eq:1.2.2}
	(\omega_{1},\omega_{2})_{H(E)}=\frac{1}{2}\sum\limits_{e\in E}\omega_{1}(e)\overline{\omega_{2}(e)}=\sum\limits_{e\in E^o}\omega_{1}(e)\overline{\omega_{2}(e)},
\end{align}
where the second equality follows from the formula $\omega_{1}(e)\overline{\omega_{2}(e)}=\omega_{1}(\be)\overline{\omega_{2}(\be)}$.

We now pass to the construction of difference operators.
Let $\nabla$ stand for the operator $\nabla:H(V)\longmapsto H(E)$, acting by the rule
\begin{align}\label{eq:1.2.3}
	(\nabla u)(e)=u(t(e))-u(o(e)), 
\end{align}
where $u\in H(V)$,  $e \in E$.
We introduce the operator $\mathrm{div}:H(E)\longmapsto H(V)$, acting by the rule
\begin{align}\label{eq:1.2.4}
	(\mathrm{div}\;\omega) (x) = \sum \limits_{e\in E(x)} \omega(e),
\end{align}
where $\omega\in H(E)$, $x\in V$.
A direct computation yields the following proposition.
\begin{prop}\label{pr:1.2.3}
	The formulas \eqref{eq:1.2.3} and \eqref{eq:1.2.4} correctly define the linear bounded operators $\nabla:H(V)\longmapsto H(E)$ and $\mathrm{div}:H(E)\longmapsto H(V)$, and moreover
	\begin{align}\label{eq:1.2.5}
		||\nabla||_{H(V)\rightarrow H(E)}\leqslant \sqrt{2\deg(G)},\quad||\mathrm{div}||_{H(E)\rightarrow H(V)}\leqslant \sqrt{2\deg(G)}.
	\end{align}
	where the number $\deg(G)$ is given by the formula \eqref{eq:1.1.1}.
	Moreover, for any $u\in H(V)$ and $\omega\in H(E)$ the following identity holds:
	\begin{align}\label{eq:1.2.6}
		(\mathrm{div}\;\omega,u)_{H(V)}=-(\omega,\nabla u)_{H(E)}.
	\end{align}
\end{prop}

In addition to the difference analogues of the gradient and divergence, we introduce multiplication operators on vertices and edges. A \textit{potential} is a bounded function $q$ from $V$ to $\mathbb{C}$. A \textit{metric} is a bounded function $a$ from $E$ to $\mathbb{C}$, even under change of arrows orientation: $a(-e)=a(e)$, $e \in E$. The corresponding multiplication operators $q:H(V)\rightarrow H(V)$ and $a:H(E)\rightarrow H(E)$ are bounded.

Our main interest will be operators of the form $-\mathrm{div}\nabla+q$ and $-\mathrm{div}a\nabla$, which we shall call the \textit{Schr\"odinger operator} and the \textit{acoustic operator}, respectively. We emphasize that, by default, we do not require positivity of the function $a$, in contrast to the differential acoustic equation, where the coefficient $a$ is usually assumed to be a positive definite matrix-valued function.
The following proposition follows from the above.
\begin{prop}\label{pr:1.2.4}
	The Schr\"odinger and acoustic operators act from $H(V)$ to $H(V)$, are linear and bounded. If the functions $a$ and $q$ are real-valued, then the operators are self-adjoint.
\end{prop}

We shall consider analogues of boundary-value problems for difference operators on a graph. Therefore we shall assume that every vertex of the graph $G=(V,E)$ is either boundary or interior. Denote by $\partial V$ the set of boundary vertices and by $\mathring{V}$ the set of interior vertices. Then $V=\mathring{V}\cup\partial V$, $\mathring{V}\cap\partial V=\varnothing$ and $\mathring{V}\neq\varnothing$; the case $\partial V=\varnothing$ is not excluded. If $\partial V \neq \varnothing$, we shall say that $G$ is a \textit{graph with boundary} $\partial V$. 
The operators from $H(V)$ to $H(\partial V)$ defined by the formulas
\begin{align}
	\label{eq:1.3.2}  (B^{D}u)(x)&=u(x),\quad x\in\partial V, \\
	\label{eq:1.3.3}  (B^{N}u)(x)&=(-\mathrm{div}a\nabla u)(x),\quad x\in\partial V,\\
	\label{eq:1.3.4}  (B^{R}u)(x)&=(-\mathrm{div}a\nabla u+ru)(x),\quad x\in\partial V,
\end{align}
are called the \textit{Dirichlet, Neumann, and Robin boundary operators} (boundary condition of the third kind), respectively. Here $r$ is a bounded function from $\partial V$ to $\bC$. We emphasize that in the formulas \eqref{eq:1.3.3} and \eqref{eq:1.3.4} the expressions $(-\mathrm{div}a\nabla u)(x)$ to the right are first computed on all of $V$ and then restricted to $\partial V$.

Consider the problem
\begin{align}\label{eq:1.3.1}
	\left\{\begin{array}{ll}
		(Au)(x)-\mu u(x)=f(x),&x\in \mathring V,\\
		(Bu)(x)=g(x),&x\in\partial V,
	\end{array}\right.
\end{align}
where $\mu \in \bR$, $A$ is an operator of the form $-\mathrm{div}a\nabla+q$, $B$ is one of the operators \eqref{eq:1.3.2}-\eqref{eq:1.3.4}, and the functions $f$ and $g$ are given functions on $\mathring V$ and $\partial V$. In the case $\partial V=\varnothing$, no boundary conditions arise, and the subsequent can be naturally simplified. From \eqref{eq:1.2.6} it follows that for any functions $u,v\in H(V)$ Green's formula holds:
\begin{align}\label{eq:1.3.5}
	(Au,v)_{H(\mathring{V})}+(B^{R}u,B^{D}v)_{H(\partial V)}=(u,A^{*}v)_{H(\mathring{V})}+(B^{D}u,(B^{R})^{*}v)_{H(\partial V)},
\end{align}
where $A^{*}=-\mathrm{div}\bar{a}\nabla+\bar{q}$ and $(B^{R})^{*}=-\mathrm{div}\bar{a}\nabla+\bar{r}$. We also note that the action of the operator $A$ can be described explicitly without using $\nabla$ and $\rdiv$:
\begin{align}\label{eq_A_explicit}
	(A u)(x)=\sum\limits_{e\in E(x)}a(e)(u(x)-u(t(e)))+q(x)u(x).
\end{align}

\subsection{Cylinder}\label{ss_Cyl}
\begin{figure}[h!]\centering
	\includegraphics[width=0.9\textwidth]{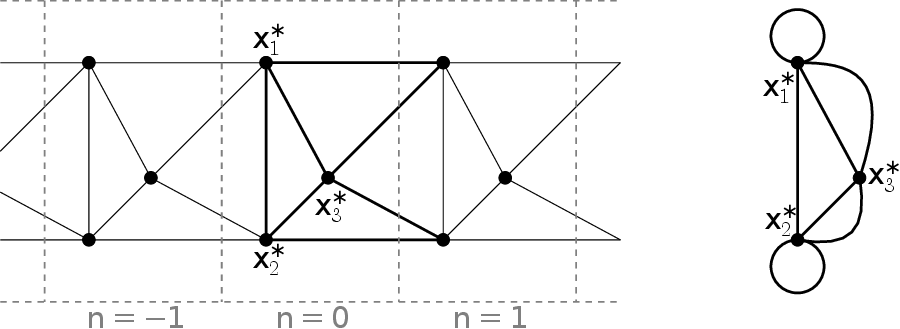}
	\caption{Example of a discrete cylinder (left) and its fundamental graph (right)}\label{fCyl}
\end{figure}

To define a cylindrical graph, we start with an intuitive picture of it. Assume that the graph is embedded in three-dimensional space. The vertices of the graph are identified with points of $\mathbb{R}^{3}$, and its edges are identified with curves in $\mathbb{R}^{3}$ joining the points corresponding to the endpoints of the edge. The dimension 3 is chosen for visual clarity, since in this case one may assume that edges can intersect only at vertices. 
Then the graph $G=(V,E)$ is called periodic with period $\vec{a}\in\mathbb{R}^{3}$ if the sets $V$ and $E$ are invariant under a shift by the vector $\vec{a}$. An example of such a graph is shown in Figure \ref{fCyl}.

Let us give a more formal definition that does not require an embedding into Euclidean space.
A pair $\trans=(\trans_{V},\trans_{E})$ is called a \textit{periodic shift} on the graph $G=(V,E)$ if $\trans$ is an isomorphism from $G$ to $G$ having no finite invariant subsets; the graph $G$ is then called \textit{periodic}. The absence of finite invariant subsets is equivalent to $\trans_{V}^{n}x\neq x$, $\trans_{E}^{n}e\neq e$ for all $n\in\mathbb{Z}, x\in V, e\in E$. If we consider the graph $G$ to have a boundary $\partial V$, then we also require $\trans_V (\partial V)=\partial V$.

We shall regard vertices $x_{1}$ and $x_{2}$ as equivalent, $x_{1}\sim x_{2}$, if $x_{1}=\trans_{V}^{n}x_{2}$ for some $n\in\mathbb{Z}$. Arrows $e_{1}$ and $e_{2}$ are called equivalent, $e_{1}\sim e_{2}$, if $e_{1}=\trans_E^{m}e_{2}$ for some $m\in\mathbb{Z}$. Then the quotient set $G^{*}=(V^{*}, E^{*})$, where $V^{*}={V}/_{\sim}$ and $E^{*}={E}/_{\sim}$, can be regarded as an ordinary graph. Indeed, we shall say that an arrow $e^{*}\in E^{*}$ leaves a vertex $x^{*}\in V^{*}$, and write $x^*=o(e^*)$, if there are representatives $x \in x^*$ and $e \in e^*$ such that $x=o(e)$; the equality $x^*=t(e^*)$ is defined analogously. Note that the graph $G^{*}$ may contain loops even if $G$ had no loops. Indeed, an arrow $e \in E$ joining vertices from the same equivalence class corresponds to a loop $e^* \in E^*$. The set $\partial V^{*}={\partial V}/_{\sim}$ is called the boundary of the graph $G^{*}$; $\partial V^{*}\subset V^{*}$.

The pair $\Pi=(G,\trans)$ is called a \textit{discrete cylinder}, or simply a cylinder, if $G$ is a periodic graph with shift operator $\trans$ and finite graph $G^{*}$ (the sets $V^{*}$ and $E^{*}$ are finite), which we shall call the \textit{fundamental graph}.
The cylinders $\Pi_{1}=(G_{1},\trans_{1})$ and $\Pi_{2}=(G_{2},\trans_{2})$ are called isomorphic if there exists an isomorphism $\varphi=(\varphi_V, \varphi_E):G_{1} \to G_{2}$ satisfying the relations
\begin{equation}\label{eq:2.1.1}
	\varphi_{V}\circ \trans_{V_{1}}=\trans_{V_{2}}\circ \varphi_{V},\quad \varphi_E\circ \trans_{E_{1}}=\trans_{E_{2}}\circ \varphi_E.
\end{equation}

Let us discuss the problem of reconstructing the cylinder $\Pi=(G,\trans)$ from its fundamental graph $G^{*}$. For this purpose we introduce the notion of the \textit{edge index}. From each equivalence class of the set $V^{*}$ choose one representative, and call the resulting set $V_{0}$ a \textit{periodicity cell}. Define a function $n(x)$ on the vertices of the graph $G$ as follows: for any $x\in V_{0}$ set $n(x)=0$, and $n(y)=m$ if $y=\trans_{V}^{m}x$. Then
\begin{equation}\label{eq:2.1.2}
	V=\bigcup\limits_{m\in\mathbb{Z}} V_{m},\;\text{where}\;V_{m}=\{x\in V:n(x)=m\};
\end{equation}
we call $V_{m}$ the $m$th periodicity cell. The function $\tau(e)=(\nabla n)(e)$ is called the edge index. Note that $\tau$ is a periodic odd function on $E$, and therefore it defines an odd function on $E^{*}$, which we shall denote by the same letter.

\begin{prop}\label{pr:2.1.4}
	Let $G^{*}$ be the fundamental graph of the cylinder $\Pi$, and let $\tau$ be the edge index. Then the cylinder $\Pi$ is uniquely reconstructed from the pair $(G^{*},\tau)$ up to isomorphism.
\end{prop}

\begin{proof}
	Set $\widetilde{V}=V^{*}\times\mathbb{Z}$. Let us join the vertices in $\widetilde{V}$ by arrows. For each arrow $e^{*} \in E^*$ and arbitrary $k \in \bZ$, we draw an arrow from the vertex $(o(e^{*}),k)$ to the vertex $(t(e^{*}),k+\tau(e^*))$ and denote the resulting arrow by $(e^{*},k)$. To this arrow there corresponds the arrow $(-e^*,k+\tau(e^*))$, which joins the same pair of vertices and have the opposite orientation; hence $-(e^{*},k):=(-e^*,k+\tau(e^*))$. On the resulting graph $\widetilde{G}=(\widetilde{V}, \widetilde{E})$ let us introduce the shift operators $\trans_{\widetilde{V}}(x^{*},k)=(x^{*},k+1)$ and $\trans_{\widetilde{E}}(e^{*},k)=(e^{*},k+1)$. It can be directly verified that they satisfy all the requirements in the definition of a shift operator. 
	Thus,  $\widetilde{G}=(\widetilde{V},\widetilde{E})$ and $\widetilde{\trans}=(\trans_{\widetilde{V}},\trans_{\widetilde{E}})$ define a cylinder $\widetilde{\Pi}$ with the fundamental graph $G^{*}$ and the edge index $\tau$ with respect to the set $\widetilde{V}_{0}=V^{*}\times\{0\}$. It remains to check that the cylinders $\Pi$ and $\widetilde{\Pi}$ are isomorphic.
	
	For all $x^{*}\in V^{*}$ set $\varphi_{V}(x^{*},0)=x$, where $x$ is a representative of the class $x^{*}$ belonging to the periodicity cell $V_{0}$. Let us extend $\varphi_{V}$ to the whole set $\widetilde{V}$: $\varphi_{V}(x^{*},k)=\varphi_{V}(\trans_{\widetilde{V}}^{k}(x^{*},0))=\trans_{V}^{k}x$. Let $e^{*}$ be an arbitrary arrow in $E^*$. Among all representatives $e$ of the class $e^{*}$, we choose the one with $o(e)=\varphi_{V}(o(x^{*}),0)$ and set $\varphi_{E}(e^{*},k)=\trans_{E}^{k}e$. By construction, it is clear that $\varphi_{V}$, $\varphi_{E}$ are bijections satisfying the formulas \eqref{eq_morph} and \eqref{eq:2.1.1}.
\end{proof}

From now on we shall use the numbering of vertices and arrows in the cylinder introduced in the proof of Proposition \ref{pr:2.1.4}, identifying $G$ and $\widetilde{G}$. Namely, for any $x \in V$ and $e \in E$ we shall write
\begin{align}
	x&=(x^*, n), \quad \text{where } x^*=[x]\in V^*, \,n=n(x)\in\bZ, \label{eq_coord_V}\\
	e&=(e^*, n), \quad \text{where } e^*=[e]\in E^*, \,n=n(o(e))\in\bZ. \label{eq_coord_E}
\end{align}
In this case, for any $e^*\in E^*$ and $n \in \bZ$ the equality
\begin{align}\label{eq_coord_E-}
	-(e^{*},n):=(-e^*,n+\tau(e^*)).
\end{align}
holds. With this numbering of vertices and arrows the following equalities hold:
\begin{align}
	o(e^*,n)&=(o(e^*),n), \quad t(e^*,n)=(t(e^*), n+\tau(e^*)), \label{eq_oe_Cyl}\\
	E(x^*,n)&=\{(e^*,n): e^*\in E^*(x^*)\}.\label{eq_E(x)_Cyl}
\end{align}
In other words, the arrows $e$ leaving the vertex $(x^*, n)$ have the form $e=(e^*,n)$, where $e^*\in E^*(x^*)$ (see \eqref{eq_E(x)}).

Let us examine how the freedom in choosing the set $V_{0}$ affects the pair $(G^{*},\tau)$. Clearly, the fundamental graph $G^{*}$ does not depend on $V_{0}$. We now determine how the choice of $V_{0}$ affects the function $\tau$. Let $V_{0}^{1}$ and $V_{0}^{2}$ be two different periodicity cells. Then the corresponding functions are $\tau_{1}=\nabla n_{1}$ and $\tau_{2}=\nabla n_{2}$. Note that
$$
\tau_{1}(e)-\tau_{2}(e)=(\nabla n)(e),
$$
where the function $n=n_{1}-n_{2}$ is defined on the vertices of the cylinder $\Pi$ and takes integer values. We show that it is periodic, and hence is well defined on $V^{*}$. Indeed, let $x_{1}\in V_{0}^{1},\;x_{2}\in V_{0}^{2}: [x_{1}]=[x_{2}]$. For any $x\in [x_{1}]$ consider $y=\trans_{V}^{m}x$. Then $x=\trans_{V}^{k}x_{1}=\trans_{V}^{l}x_{2}$, $y=\trans_{V}^{k'}x_{1}=\trans_{V}^{l'}x_{2}$, where $l'-l=k'-k=m$. We have
$$
n_{1}(x)-n_{2}(x)=k-l=k'-l'+(k-k')-(l-l')=k'-l'+m-m=k'-l'=n_{1}(y)-n_{2}(y).
$$
For any periodicity cell $V_0^{1}$, with a suitable choice of $V_0^{2}$, one can obtain as $n$ an arbitrary integer-valued function on $V^{*}$. This means that adding $\nabla n$, $n :V^*\to \bZ$, to the function $\tau$ from Proposition \ref{pr:2.1.4} does not change the cylinder.

By varying integer-valued odd functions $\tau$ on $E^*$, we obtain all possible cylinders $(G,\trans)$ with the fundamental graph $G^*$. However, in some cases the resulting graph $G$ may be disconnected even for a connected graph $G^*$. We describe the conditions under which this happens. Let $M$ be a finite set of integers. Let $\gcd(M)$ stand for the greatest common divisor of all elements in $M$, assuming that $\gcd(m_{1},\ldots,m_{k})=\gcd(|m_{1}|,\ldots,|m_{k}|)$ and $\gcd(0,m_{1},\ldots,m_{k})=\gcd(m_{1},\ldots,m_{k})$ for any $m_{1},\ldots,m_{k}\in\mathbb{Z}$.
\begin{prop}\label{p_Cyl_conn}
	Let $G^*=(V^*,E^*)$ be a fundamental graph with index function $\tau$. Let also $G_0^*=(V^*, E_0^*)$ be a spanning tree of the graph $G^*$. Then\\
	{\rm1.} the function $n:V^* \to \bZ$ can be chosen so that $\tau_0=\tau+\nabla n$ satisfies the equalities $\tau_0(e^*)=0$ for all $e^* \in E_0^*$;\\
	{\rm2.} the graph $G$ corresponding to the pair $(G^*, \tau)$ is connected if and only if $\gcd(\tau_0(E^*))=1$, where $\tau_0(E^*)=\{\tau_0(e^*), e^*\in E^*\}$;\\
	{\rm3.} if $\gcd(\tau_0(E^*))=d$, then the graph $G$ consists of $d$ connected components, each isomorphic to the cylinder with the fundamental graph $G^*$ and the index function $\tau_0/d$.
\end{prop}
\begin{proof}
	1. Since $G_0^*$ is a tree, for any pair of vertices $x_0^*, x^* \in V^*$ there exists a unique geodesic $\gamma(x_0^*,x^*)$ in $G_{0}^{*}$ with initial point $x_0^*$ and terminal point $x^*$. We regard the vertex $x_0^* \in V^*$ as fixed and, for an arbitrary vertex $x^* \in V^*$, set 
	\begin{align}
		n(x^*)=-\sum_{j=1}^{m}\tau(e_j^*),
	\end{align}
	where $(e_1^*, \dots, e_m^*)=\gamma(x_0^*,x^*)$. Let now $e^*$ be an arbitrary arrow in $E_0^*$. Then either $e^* \in \gamma(x_0^*, t(e^*))$, or $-e^* \in \gamma(x_0^*, o(e^*))$. In the first case we have $\gamma(x_0^*, t(e^*))=(e_1^*, \dots, e_{m-1}^*, e^*)$ and $\gamma(x_0^*, o(e^*))=(e_1^*, \dots, e_{m-1}^*)$, hence
	\begin{align}
		(\nabla n)(e^*)=n(t(e^*))-n(o(e^*))=-\tau(e^*).
	\end{align}
	The equality $(\nabla n)(e^*)=-\tau(e^*)$ is verified similarly in the second case. Thus the function $\tau_0=\tau+\nabla n$ is zero on the set $E_0^*$, and item 1 is proved.
	
	2. We apply the construction of a discrete cylinder from the proof of Proposition \ref{pr:2.1.4}, choosing $\tau_0$ as the index function. The vertex set $V$ is a union of the form \eqref{eq:2.1.2} of periodicity cells $V_m=\{(x^*, m), x^* \in V^*\}$. An arrow in the set 
	$$E_0^*\times \bZ= \{(e^*, m), \,e^* \in E_0^*,\, l\in \bZ\}$$ 
	 join vertices inside the single cell $V_m$, since  $\tau_0(e^*)=0$. By assumption, $G_0^*$ is a spanning tree of the graph $G^*$, then each cell $V_m$, $m \in \bZ$, joined by such arrows, becomes connected.
	It remains to determine under what condition there exists a geodesic joining the cell $V_0$ to an arbitrary cell $V_m$, which is equivalent to the connectedness of the graph $G$. 
	
	Let $\gamma(x_0, x)=(e_1, \dots, e_k)$ be a geodesic in $G$ joining the points $x_0   \in V_0$ and $x \in   V_m$, i.e., $o(e_1)=x_0$, $t(e_k)=x$ and $o(e_{j+1})=t(e_j)$, $j=1, \dots, k-1$. We have
	\begin{align}\label{eq_shift_geo}
		m=n(x)-n(x_0)=\sum_{j=1}^k \big(n(t(e_j))-n(o(e_j))\big)=\sum_{j=1}^{k} \tau_0(e_j).
	\end{align}
	Here $n(x)$ stands for the number of the cell $V_{n(x)}$ containing $x$. For each edge of $G^*$ we choose an orientation and denote the obtained arrows by $\varepsilon_1^*, \dots, \varepsilon_{K}^*$, $K=|E^*|/2$. Then, by the periodicity of the function $\tau_0$, equality \eqref{eq_shift_geo} takes the form
	\begin{align}\label{eq_bezout}
		m=\sum_{j=1}^{K}q_j \tau_0(\varepsilon_j^*)
	\end{align}
	with coefficients $q_j \in \bZ$. By B\'ezout's identity, this equality is solvable if and only if the number $m$ is divisible by $\gcd(\tau(\varepsilon_1^*), \ldots, \tau(\varepsilon_{K}^*))$. 
	
	If the graph $G$ is connected, then there exists a geodesic joining the cell $V_0$ to the cell $V_1$; hence $\gcd(\tau(\varepsilon_1^*), \ldots, \tau(\varepsilon_{K}^*))=1$. Now assume that $\gcd(\tau(\varepsilon_1^*), \ldots, \tau(\varepsilon_{K}^*))=1$, then formula \eqref{eq_bezout} holds with $m=1$. The numbers $q_j$ can be made non-negative by changing $\varepsilon_j^*$ for $-\varepsilon_j^*$, if necessary. 
	Let us describe the construction of a geodesic connecting $V_0$ to $V_1$. Recall that arrows in $E_0^*\times \bZ$ 
	join vertices inside a single cell and can therefore be added without any restrictions.  Let us set $n_1=0$, $n_{j+1}=n_{j}+q_j \tau_0(\varepsilon_j^*)$, $j=1, \dots K$. Then $n_{K+1}=1$. Now the required geodesic 
	consists of the arrows $(\varepsilon_j^*,n_j+k\tau_0(\varepsilon_j^*))$, $k=0, \dots, q_j-1$, $j=1, \dots, K$, connected by arrows in $E_0^*\times \bZ$.
 Finally, if $\gcd(\tau(\varepsilon_1^*), \ldots, \tau(\varepsilon_{K}^*))=d$, then one can construct a geodesic joining the cell $V_m$ to the cell $V_{m+d}$. Then the graph $G$ splits into $d$ isomorphic connected components.
\end{proof}
A discrete cylinder is a special case of a topological crystal, which was studied in detail in the monograph \cite{Su13}. Therefore,  in this subsection only Proposition \ref{p_Cyl_conn} appears to be novel. Compared to \cite{Su13}, we employ a different terminology and different notations that were used in the paper \cite{KorSab1}.

\subsection{Fourier transform in the cylinder}
The aim of this section is to introduce an analogue of the Fourier transform for functions $u \in H(V)$ and $\omega \in H(E)$ along the ``axis'' of the cylinder $\Pi$. Thus, taking into account the formula \eqref{eq_coord_V}, for a function $u:V \to \bC$ with finite support we set
\begin{align}
	\hat{u}(x^{*},\lambda)=\frac{1}{\sqrt{2\pi}}\sum\limits_{n\in\mathbb{Z}}\exp(-in\lambda)\;u(x^{*},n) \label{eq_fourier_V}
\end{align}
for each $x^* \in V^*$. The formula 
\begin{align}
	u(x^{*},n)=\frac{1}{\sqrt{2\pi}}\int\limits_{0}^{2\pi}\exp(in\lambda)\;\hat{u}(x^{*},\lambda)\mathrm{d}\lambda, \quad x^* \in V^*,  \label{eq_fourier_V_inv}
\end{align}
defines the inverse transform. For a fixed $x^*$, these formulas coincide with the Fourier transform of a periodic function (the Fourier series). In the literature, one can encounter other names for the transformation of type \eqref{eq_fourier_V}: the Gelfand transform or the Floquet–Bloch transform. 

The operator $\cF_V: u \mapsto \hat{u}$ extends to the unitary operator 
$\cF_V: H(V) \to \widehat{H}(V^{*}, \bT)$.
Here $\bT=\bR/{2\pi \bZ}$ is the unit circle, and $\widehat{H}(V^{*}, \bT)=L_{2}(\bT;H(V^{*}))$ is the space of functions $\hat{u}:V^*\times \bR\to \bC$, $2\pi$-periodic in the variable $\lambda$, with the inner product 
\begin{align}\label{eq_fourier_V_dualspace}
	(\hat{u}_{1},\hat{u}_{2})_{\widehat{H}(V^{*}, \bT)}=\int\limits_{0}^{2\pi}\sum\limits_{x^{*}\in V^{*}}\hat{u}_{1}(x^{*},\lambda)\overline{\hat{u}_{2}(x^{*},\lambda)}\mathrm{d}\lambda.
\end{align}
We now turn to the discussion of the Fourier transform for functions $\omega \in H(E)$. Such a function is odd: $\omega(-e)=-\omega(e)$ for every arrow $e \in E$. By \eqref{eq_coord_E-}, in the coordinates \eqref{eq_coord_E} this equality takes the form
\begin{equation}\label{eq:2.2.3}
	\omega(-e^{*},n)=-\omega(e^{*},n-\tau(e^{*})).
\end{equation}
For a function $\omega \in H(E)$ with finite support we set
\begin{align}\label{eq_fourier_E}
	\hat{\omega}(e^{*},\lambda)=\frac{1}{\sqrt{2\pi}}\sum\limits_{n\in\mathbb{Z}}\exp(-in\lambda)\;\omega(e^{*},n).
\end{align}
For all $\lambda$, the equality 
\begin{align}\label{eq_fourier_E-}
	\hat{\omega}(-e^{*},\lambda)=-\exp\left({-i\tau(e^{*})\lambda}\right)\hat{\omega}(e^*,\lambda)
\end{align} 
holds. Indeed, by \eqref{eq:2.2.3} and \eqref{eq_fourier_E} we have
\begin{align*}
	\hat{\omega}(-e^{*},\lambda)&=\frac{1}{\sqrt{2\pi}}\sum\limits_{n\in\mathbb{Z}}\exp(-in\lambda)\;\omega(-e^{*},n)=-\frac{1}{\sqrt{2\pi}}\sum\limits_{n\in\mathbb{Z}}\exp(-in\lambda)\;\omega(e^{*},n-\tau(e^{*}))=\nonumber\\
	&=-\frac{\exp\left({-i\tau(e^{*})\lambda}\right)}{\sqrt{2\pi}}\sum\limits_{m\in\mathbb{Z}}\exp(-im\lambda)\;\omega(e^{*},n)=-\exp\left({-i\tau(e^{*})\lambda}\right)\hat{\omega}(e^*,\lambda), 
\end{align*}
where $m=n-\tau(e^*)$.
The inverse transform is given by the formula
\begin{align}\label{eq_fourier_E_inv}
	\omega(e^{*},n)=\frac{1}{\sqrt{2\pi}}\int\limits_{0}^{2\pi}\exp(in\lambda)\;\hat{\omega}(e^{*},\lambda)\mathrm{d}\lambda.
\end{align}
The operator $\cF_E: \omega \mapsto \hat{\omega}$ extends to the unitary operator 
$\cF_E: H(E) \to \widehat{H}(E^{*}, \bT)$,
where $\widehat{H}(E^{*}, \bT)$ is the space of functions $\hat{\omega}:E^*\times \bR\to \bC$, $2\pi$-periodic in the variable $\lambda$ and satisfying \eqref{eq_fourier_E-}, with the inner product
\begin{align}\label{eq_fourier_E_dualspace}
	(\hat{\omega}_{1},\hat{\omega}_{2})_{\widehat{H}(E^{*}, \bT)}=\frac{1}{2}\int\limits_{0}^{2\pi}\sum\limits_{e^{*}\in E^{*}}\hat{\omega}_{1}(e^{*},\lambda)\overline{\hat{\omega}_{2}(e^{*},\lambda)}\mathrm{d}\lambda.
\end{align}

	\section{Problem in a cylinder with periodic coefficients}\label{s_ProbCyl}
Let $\Pi=(G,\cT)$ be a cylinder, where $G=(V,E)$ is a periodic loopless graph with period $\cT$ and boundary $\partial V \subset V$. Consider the problem
\begin{equation}\label{eq:2.2.1}
	\left\{\begin{array}{ll}
		(Au)(x)-\mu u(x)=f(x),&x\in\mathring{V},\\
		(Bu)(x)=g(x),&x\in\partial V.
	\end{array}\right.
\end{equation}
Here $A$ is an operator of the form $-\mathrm{div}a\nabla+q$, $B$ is a boundary operator of Dirichlet, Neumann, or Robin type (see \eqref{eq:1.3.2}-\eqref{eq:1.3.4}), $\mu \in \bR$ is a spectral parameter, $\mathring{V}=V\setminus\partial V$, and $f$ and $g$ are given functions on $\mathring{V}$ and $\partial V$, respectively. We assume that the ``metric'' $a$ is a real-valued function on $E$, even, $a(-e)=a(e)$, and periodic with the period $\cT_{E}$: $a(\cT_{E}(e))=a(e)$. The ``potentials'' $q$ and $r$ ( the latter is introduced in the case of the Robin boundary condition \eqref{eq:1.3.4}) are real-valued functions on $\mathring{V}$ and $\partial V$, respectively, periodic with the period $\cT_{V}$: $q(\cT_{V}(x))=q(x),\;r(\cT_{V}(x))=r(x)$. Throughout, we assume that the graph $G$ is connected; otherwise it splits into $k$ connected components, and the problem \eqref{eq:2.2.1} decomposes into $k$ independent problems.
In this paper we restrict ourselves to considering the problem without boundary conditions or the Dirichlet problem. Using the coordinates $x=(x^*,n)$ in $V=V^*\times \bZ$ from \eqref{eq_coord_V}, we write formula \eqref{eq_A_explicit} for the operator $A$ in the form
\begin{align}\label{eq_A_explicit_Cyl}
	\big(A u\big)(x^*,n)=\sum\limits_{e^*\in E^*(x^*)}a(e^*)\Big(u\big(x^*,n\big)-u\big(t(e^*), n+\tau(e^*)\big)\Big)+q(x^*)u(x^*,n).
\end{align}
Here we have taken into account the periodicity of the functions $a$ and $q$, as well as formulas \eqref{eq_oe_Cyl} and \eqref{eq_E(x)_Cyl}.
\subsection{Problem on the fundamental graph.}
Applying the Fourier transform $\cF_V$ to the left- and right-hand sides of the equations \eqref{eq:2.2.1}, we obtain the problem
\begin{equation}\label{eq:2.2.16}
	\left\{\begin{array}{ll}
		(A(\lambda)\hat{u})(x^{*},\lambda)-\mu \hat{u}(x^{*},\lambda)=\hat{f}(x^{*},\lambda),&x^{*}\in\mathring{V}^{*}\\
		(B(\lambda)\hat{u})(x^{*},\lambda)=\hat{g}(x^{*},\lambda),&x^{*}\in\partial V^{*}
	\end{array}\right.
\end{equation}
on the fundamental graph $G^{*}$ with parameter $\lambda \in \bT$.

To describe explicitly the action of the operator $A(\lambda)$, we introduce the operators $\nabla(\lambda)$ and $\rdiv(\lambda)$. For an arbitrary function $v: V^* \to \bC$ and a value $\lambda \in \bC$, we consider on $V$ the function $w(x^*,n)=\exp\left(i n \lambda\right)v(x^*)$ and apply to it the operator $\nabla$ from \eqref{eq:1.2.3}:
\begin{align}\label{eq_nabla_exp}
	(\nabla w)(e^*,n)=w\big(t(e^*),n+\tau(e^*)\big)-w\big(o(e^*),n\big)=\exp(i n \lambda)\big(\nabla(\lambda)v\big)(e^*),
\end{align}
where $\nabla(\lambda)v$ is a function from $E^*$ to $\bC$, given by the formula
\begin{align}\label{eq_nabla_lam}
	\big(\nabla(\lambda)v\big)(e^*):=\exp\left(i \tau(e^*) \lambda\right)v\big(t (e^*)\big)-v\big(o(e^*)\big).
\end{align}
When $e^*$ is replaced by $-e^*$, this function satisfies the conditions \eqref{eq_fourier_E-}:
\begin{align*}
	\big(\nabla(\lambda)v\big)(-e^*)&=\exp\left(i \tau(-e^*) \lambda\right)v\big(t(-e^*)\big)-v\big(o(-e^*)\big)=\nonumber\\
	&=-\exp\left(-i \tau(e^*) \lambda\right)\left(\exp\left(i \tau(e^*) \lambda\right)v\big(t (e^*)\big)-v\big(o(e^*)\big)\right)=\nonumber\\
	&=-\exp\left(-i \tau(e^*) \lambda\right)\big(\nabla(\lambda)v\big)(e^*).
\end{align*}
Similarly, for an arbitrary function $\varphi: E^* \to \bC$ and a number $\lambda \in \bC$, we consider on $E$ the function $\psi(e^*,n)=\exp\left(i n \lambda\right)\varphi(e^*)$ and apply to it the operator $\rdiv$ from \eqref{eq:1.2.4}; taking into account \eqref{eq_E(x)_Cyl}, we have
\begin{align*}
	(\rdiv\psi) (x^*,n) = \sum \limits_{e^*\in E^*(x^*)} \psi(e^*,n)=\exp\left(i n \lambda\right)(\rdiv(\lambda)\varphi) (x^*),
\end{align*}
where $\rdiv(\lambda)\varphi$ is a function from $V^*$ to $\bC$, given by the expression
\begin{align}\label{eq_div_exp}
	(\rdiv(\lambda)\varphi) (x^*):=\sum \limits_{e^*\in E^*(x^*)} \varphi(e^*).
\end{align}

Let $\lambda\in \bC$ be fixed, $v: V^* \to \bC$ be an arbitrary function, and the function $\varphi:E^* \to \bC$ satisfy a condition of the form \eqref{eq_fourier_E-}:
$
\varphi(-e^{*})=-\exp\left({-i\tau(e^{*})\lambda}\right)\varphi(e^*).
$
The identity
\begin{equation}\label{eq:2.2.15}
	\big(\mathrm{div}(\lambda)\varphi,v\big)_{H(V^{*})}=-\big(\varphi,\nabla(\bar{\lambda})v\big)_{H(E^{*})}.
\end{equation}
can be verified by a direct computation.
We emphasize that the ordinary inner products \eqref{eq:1.2.1} and \eqref{eq:1.2.2} are involved here, without integration over $\lambda$. 
\begin{prop}
	The following relations hold
	\begin{align}\label{eq_nabla_fourier}
		\cF_E \nabla=\nabla(\lambda)\cF_V \quad \text{on } H(V),\qquad \cF_V \rdiv=\rdiv(\lambda)\cF_E \quad \text{on } H(E).
	\end{align}
\end{prop}
\begin{proof}
	We begin with the verification of the first identity in \eqref{eq_nabla_fourier}. For a function $u \in H(V)$ with finite support, we derive from formulas \eqref{eq_fourier_V} and \eqref{eq_nabla_exp}
	\begin{align*}
		(\nabla u)(e^{*},n)=\frac{1}{\sqrt{2\pi}}\int\limits_{0}^{2\pi}\nabla \big(\exp(in\lambda)\,\hat{u}(\cdot, \lambda)\big)(e^{*},\lambda)\mathrm{d}\lambda=\frac{1}{\sqrt{2\pi}}\int\limits_{0}^{2\pi}\exp(in\lambda)\,\left(\nabla(\lambda)\hat{u}\right)(e^{*},\lambda)\mathrm{d}\lambda.
	\end{align*}
	It follows that the function $\nabla(\lambda)\hat{u}$ is the Fourier transform of the function $\nabla u$, which is equivalent to $\cF_E \nabla u=\nabla(\lambda)\cF_V u$. Since all operators in this identity are bounded, it extends to all $H(V)$. The second identity in \eqref{eq_nabla_fourier} is derived from \eqref{eq_fourier_E} and \eqref{eq_div_exp} by similar reasoning.
\end{proof}

We now turn directly to the discussion of the operator $A(\lambda)$. In our case, the metric and the potential do not depend on $n$, and therefore they correctly define the multiplication operators by $a(e^{*})$ and $q(x^{*})$ on the fundamental graph. The same is true for the potential $r(x^*)$ from the third boundary condition. Then for an arbitrary function $v: V^* \to \bC$ the action of the operator $A(\lambda)$ is given by the expression
\begin{equation}\label{eq:2.2.17}
	(A(\lambda)v)(x^{*})=(-\mathrm{div}(\lambda)a\nabla(\lambda)v)(x^{*})+q(x^{*})v(x^{*}), \quad x^* \in \mathring{V}^*,
\end{equation}
and the boundary operator $B(\lambda)$ is defined by one of the following formulas
\begin{align}
	(B^{D}(\lambda)v)(x^{*})&=v(x^{*}),\quad x^{*}\in\partial V^{*},\nonumber\\
	(B^{N}(\lambda)v)(x^{*})&=(-\mathrm{div}(\lambda)a\nabla(\lambda)v)(x^{*}),\quad x^{*}\in\partial V^{*},\label{eq:2.2.18}\\
	(B^{R}(\lambda)v)(x^{*})&=(-\mathrm{div}(\lambda)a\nabla(\lambda)v)(x^{*})+r(x^{*})v(x^{*}),\quad x^{*}\in\partial V^{*}.\nonumber
\end{align}
From \eqref{eq:2.2.15} the Green's formula follows (cf. \eqref{eq:1.3.5}):
\begin{align}\label{eq:2.2.21}
	(A(\lambda)v_1,v_2)_{H(\mathring{V}^{*})}&+(B^{R}(\lambda)v_1,B^{D}(\bar{\lambda})v_2)_{H(\partial V^{*})}=\nonumber\\
	&=(v_1,A(\bar{\lambda})v_2)_{H(\mathring{V}^{*})}+(B^{D}(\lambda)v_1,B^{R}(\bar{\lambda})v_2)_{H(\partial V^{*})},
\end{align}
where $v_1,\,v_2$ are arbitrary functions in $H(V^{*})$. Thus, the boundary value problem \eqref{eq:2.2.16} is self-adjoint with respect to Green's formula \eqref{eq:2.2.21} for $\lambda \in \bR$.

It is clear that in any strip $\lambda \in \{\bC: |\rIm \lambda|<\gamma\}$ the operators $\nabla(\lambda)$ and $\rdiv(\lambda)$ are uniformly bounded (cf. Proposition \ref{pr:1.2.3}), and therefore the operator $A(\lambda)$ is uniformly bounded in such a strip. By \eqref{eq_nabla_exp} and \eqref{eq_div_exp}, the identity
\begin{align}\label{eq_A_exp}
	(A w)(x^*,n)=\exp(i n \lambda)\big(A(\lambda)v\big)(x^*)
\end{align}
holds for the function $w(x^*,n)=\exp\left(i n \lambda\right)v(x^*)$.
Finally, the action of the operator $A(\lambda)$ can be described explicitly without using $\nabla(\lambda)$ and $\rdiv(\lambda)$:
\begin{align}\label{eq_A_explicit_lam}
	(A(\lambda) v)(x^*)=\sum\limits_{e^*\in E^*(x^*)}a(e^*)\big(v(x^*)-\exp(i\lambda\tau(e^*))v(t(e^*))\big)+q(x^{*})v(x^{*}).
\end{align}

\subsection{Spectrum of the problem on the fundamental graph}\label{ss_Fund_spec}
In Sections \ref{ss_Fund_spec}-\ref{ss_Cyl_waves} we assume that the boundary $\partial V$ is absent (in Section \ref{ss_Cyl_Dir} the results of these sections are extended to the Dirichlet problem). Then no boundary conditions arise, and the problem \eqref{eq:2.2.16} takes the form
\begin{equation}\label{eq:2.3.1}
	(A(\lambda)\hat{u})(x^{*},\lambda)-\mu \hat{u}(x^{*},\lambda)=\hat{f}(x^{*},\lambda),\quad x^{*}\in V^{*}.
\end{equation}
We consider \eqref{eq:2.3.1} as a problem in $H(V^{*})$ that depends on the parameter $\lambda \in \bC$. We are to study the properties of the resolvent of the operator $A(\lambda)$ and to obtain its representation near the real axis $\lambda\in \bR$. The operator $A(\lambda)$ is an analytic $2\pi$-periodic function of the parameter $\lambda\in\mathbb{C}$, and by \eqref{eq:2.2.21}, for all $\lambda\in\mathbb{R}$ this operator is self-adjoint in the space $H(V^{*})$.

The operator $A(\lambda)$ acting in the finite-dimensional space $H(V^*)$ is unitarily equivalent to a matrix  $M(\lambda)$ acting in the space $\bC^{|V^*|}$. Therefore its characteristic polynomial $\det(A(\lambda)-\mu)$ is well defined.
We shall say that the value $\mu$ corresponds to a \textit{flat band} of the problem \eqref{eq:2.2.1} if $\det(A(\lambda)-\mu)=0$ for all $\lambda\in\mathbb{C}$. Throughout this paper we assume that $\mu$ does not correspond to any flat band.

\begin{prop}\label{lm:2.3.2}
	Assume that $\mu$ does not coincide with any flat band of the problem \eqref{eq:2.2.1}. Then the operator-valued function $\lambda\mapsto R(\lambda,\mu)=(A(\lambda)-\mu)^{-1}$ is meromorphic and $2\pi$-periodic in $\mathbb{C}$, and in the vertical strip $\mathrm{Re}\;\lambda\in[0,2\pi)$ it can have at most finitely many poles.
\end{prop}

\begin{proof}
	The resolvent $R(\lambda,\mu)$ of the operator $A(\lambda)-\mu$ corresponds to the resolvent of the matrix $M(\lambda)-\mu$; in other words,
	$$
	R(\lambda,\mu)\sim (M(\lambda)-\mu)^{-1}=\frac{\mathrm{adj}(M(\lambda)-\mu)}{\det(M(\lambda)-\mu)},
	$$
	where $\mathrm{adj}(M(\lambda)-\mu)$ is the adjugate matrix, composed of the algebraic cofactors of the corresponding entries of the transposed matrix. Each entry of the matrix $M(\lambda)$ is a trigonometric polynomial in $\lambda$. Therefore the elements of the matrix $(M(\lambda)-\mu)^{-1}$ have the form
	$$
	\frac{\sum\limits_{k=-L_{1}}^{L_{1}}a_{k}e^{ik\lambda}}{\sum\limits_{k=-L_{2}}^{L_{2}}b_{k}e^{ik\lambda}},
	$$
	where $a_{k}$ and $b_{k}$ are real numbers depending only on $\mu$. Since the number $\mu$ is not a flat band, the denominator is not identically zero. Moreover, the operator $R(\lambda,\mu)$ is a rational function of the arguments $e^{i\lambda}$ and $\mu$. Hence the statement of the lemma follows obviously.
\end{proof}

We now discuss the spectrum of $A(\lambda)$. For brevity, set $N=|V^{*}|$. According to \cite{Kato}, Chapter 2, \S6, there exist functions $\{\lambda\mapsto\mu_{j}(\lambda)\in \bC\}_{j=1}^{N}$ and $\{\lambda\mapsto\psi_{j}(\lambda)\in H(V^*)\}_{j=1}^{N}$, analytic in a symmetric neighborhood $\mathbb{U}$ of the real axis, such that
\begin{align}\label{eq_A_lam_eigenfun}
	A(\lambda)\psi_{j}(\lambda)=\mu_{j}(\lambda)\psi_{j}(\lambda),\quad\forall \lambda\in\mathbb{U},
\end{align}
the vectors $\psi_{1}(\lambda),\ldots,\psi_{N}(\lambda)$ form a basis in $H(V^{*})$ and the relations
\begin{align}\label{eq:2.3.2}
	&\mu_{j}(\bar{\lambda})=\overline{\mu_{j}(\lambda)},\;\lambda\in\mathbb{U},\nonumber\\
	&(\psi_{j}(\lambda),\psi_{l}(\bar{\lambda}))_{H(V^{*})}=\delta_{jl},\;\lambda\in\mathbb{U}.
\end{align}
hold.
In particular, for real $\lambda$ the operator $A(\lambda)$ is self-adjoint in $H(V^{*})$, the numbers $\mu_{j}(\lambda)$ are real, and the vectors $\{\psi_{j}(\lambda)\}_{j=1}^{N}$ form an orthonormal basis. The functions $\mathbb{R}\ni\lambda\mapsto\mu_{j}(\lambda) \in \bR$ will be called \textit{band functions}.
In general, it is possible that $\mu_{j}(\lambda)=\mu_{k}(\lambda)$ for all $\lambda\in\mathbb{U}$ and $j\neq k$. In this case we shall say that \textit{constant degeneracy} occurs. Moreover, it is possible that $\mu_{j}(\lambda)=\mathrm{const}$ for all $\lambda\in\mathbb{U}$; this is equivalent to the existence of a flat band.

Let us study the dependence on $\mu$ of the singular part of the function $\bU\ni \lambda\mapsto R(\lambda,\mu)$:
\begin{equation}\label{eq:ResWithout}
	R(\lambda,\mu)=\sum\limits_{j=1}^{N}\frac{(\cdot,\psi_{j}(\bar{\lambda}))_{H(V^{*})}\psi_{j}(\lambda)}{\mu_{j}(\lambda)-\mu}.
\end{equation}
Note that the operators $A(\lambda)$ and $R(\lambda, \mu)$ are $2\pi$-periodic in $\lambda$, but the functions $\lambda \mapsto \mu_j(\lambda), \psi_j(\lambda)$ need not possess this property. One can prove that each of the functions $\lambda\mapsto \mu_j(\lambda)$ has a period that is a multiple of $2\pi$; we shall not use this fact and therefore leave it without proof.

We call a number $\mu_0$ a \textit{threshold} if there exist a value $\lambda_{0}\in\mathbb{R}$ and an index $j$ such that
\begin{equation}\label{eq:2.3.4}
	\mu_{j}(\lambda)-\mu_0=c(\lambda-\lambda_{0})^{\varkappa}(1+o(1))
\end{equation}
in a neighborhood of the point $\lambda_{0}$, where $\varkappa \geqslant 2$ is a natural number and $c\neq 0$. Since the operator $A(\lambda)$ is bounded, the band functions are bounded, and therefore each of them yields at least two thresholds, corresponding to the global maximum and minimum. Thresholds corresponding to local maxima and minima are also possible. If $\mu$ coincides with neither a flat band nor a threshold, then all poles of the function $\lambda\mapsto R(\lambda,\mu)$ in $\bU$ are simple.
If $\mu$ is a threshold but not a flat band, then the resolvent has a pole of the order $\varkappa$ in \eqref{eq:2.3.4}.
\subsection{Solvability of problem in cylinder. Asymptotics of solutions}\label{ss_Cyl_asym}
Recall that the problem \eqref{eq:2.2.16} depends on the parameter $\lambda \in \bT=\bR/2\pi\bZ$, i.e., the coefficients and the right-hand side of the equations \eqref{eq:2.2.16} are assumed to be $2\pi$-periodic in $\lambda$. By Lemma \ref{lm:2.3.2}, the solution $\hat{u}(x^{*},\lambda)$ will also be $2\pi$-periodic. Below, by default, we identify $\bT$ with the interval $[0, 2\pi)$.
If for $\lambda\in\mathbb{T}$ the resolvent $\lambda\mapsto R(\lambda,\mu)$ has no singularities, then the problem
\begin{equation}\label{eq:2.3.5}
	(Au)(x)-\mu u(x)=f(x),\quad x\in V,
\end{equation}
with right-hand side $f\in H(V)$ has a unique solution $u\in H(V)$:
\begin{equation}\label{eq:2.3.6}
	u(x^{*},n)=\frac{1}{\sqrt{2\pi}}\int\limits_{\mathbb{T}}\exp(in\lambda)R(\lambda,\mu)\hat{f}(x^{*},\lambda)\mathrm{d}\lambda,
\end{equation}
where $\hat{f}$ is the Fourier transform \eqref{eq_fourier_V} of the function $f$ (a rigorous proof of a more general fact will be given in Theorem \ref{th:2.3.7}). In the case where the resolvent has singularities on the interval $\mathbb{T}$, formula \eqref{eq:2.3.6} is not applicable. However, for a suitable right-hand side one can construct a solution by replacing $\mathbb{T}$ with $\mathbb{T}+i\nu:=\big(\bR+i\nu\big)/2\pi\bZ$. Moreover, this replacement turns out to be useful also in deriving asymptotic formulas for solutions.

Here and below we assume that the parameter $\mu$ is fixed, different from thresholds, and is not a flat band. Then, by Proposition \ref{lm:2.3.2}, the operator-valued function $\lambda\mapsto R(\lambda,\mu)$ is meromorphic on $\mathbb{C}$ and has a finite number of poles in the strip $\mathrm{Re}\;\lambda\in[0,2\pi)$. For any $\nu\in\mathbb{R}$ we introduce the space $H_{\nu}(V)$ of functions on $V$ with the norm
\begin{equation}\label{eq:2.3.7}
	||u||_{H_{\nu}(V)}=\left(\sum\limits_{x^{*}\in V^{*}}\sum\limits_{n\in\mathbb{Z}}\exp{(2\nu n)}|u(x^{*},n)|^{2}\right)^{1/2}.
\end{equation}
For $u\in H_{\nu}(V)$ we consider the Fourier transform
\begin{equation}\label{eq:2.3.8}
	\begin{gathered}
		\hat{u}(x^{*},\lambda)=\frac{1}{\sqrt{2\pi}}\sum\limits_{n\in\mathbb{Z}}\exp(-in\lambda)\;u(x^{*},n)
	\end{gathered}
\end{equation}
as a function of $\lambda\in\mathbb{T}+i\nu$. By Parseval's identity
\begin{equation}\label{eq:2.3.9}
	\sum\limits_{x^{*}\in V^{*}}\sum\limits_{n\in\mathbb{Z}}\exp{(2\nu n)}|u(x^{*},n)|^{2}=\sum\limits_{x^{*}\in V^{*}}\int\limits_{\mathbb{T}+i\nu}|\hat{u}(x^{*},\lambda)|^{2}\mathrm{d}\lambda,
\end{equation}
the transform \eqref{eq:2.3.8} implements an isometric isomorphism
$$
\mathcal{F}_{V}^{\nu}:H_{\nu}(V)\to \widehat{H}(V^*;\bT+i\nu):=L_{2}(\mathbb{T}+i\nu,H(V^{*})).
$$
The inverse transform is given by the formula
\begin{equation}\label{eq:2.3.10}
	u(x^{*},n)=\frac{1}{\sqrt{2\pi}}\int\limits_{\mathbb{T}+i\nu}\exp(in\lambda)\;\hat{u}(x^{*},\lambda)\mathrm{d}\lambda.
\end{equation}

\begin{thm}\label{th:2.3.7}
	Let $\nu\in\mathbb{R}$ be such a number that $\mathbb{T}+i\nu$ contains no poles of the operator-valued function $\lambda\mapsto R(\lambda,\mu)$. Then for any right-hand side $f\in H_{\nu}(V)$ there exists a unique solution $u\in H_{\nu}(V)$ of the problem \eqref{eq:2.3.5}. Such a solution is given by the formula
	\begin{equation}\label{eq:2.3.11}
		u(x^{*},n)=\frac{1}{\sqrt{2\pi}}\int\limits_{\mathbb{T}+i\nu}\exp(in\lambda)R(\lambda,\mu)\hat{f}(x^{*},\lambda)\mathrm{d}\lambda,
	\end{equation}
	and the estimate
	\begin{equation}\label{eq:2.3.12}
		||u||_{H_{\nu}(V)}\leqslant\mathrm{const}||f||_{H_{\nu}(V)}.
	\end{equation}
	holds.
\end{thm}

\begin{proof}
	The transform \eqref{eq:2.3.8} reduces the problem \eqref{eq:2.3.5} to the equation \eqref{eq:2.3.1} for almost all $\lambda\in\mathbb{T}+i\nu$. Its solution is given by the function  $\hat{u}(\cdot,\lambda)=(R(\lambda,\mu)\hat{f})(\cdot,\lambda)$. Since $\lambda\mapsto R(\lambda,\mu)$ is uniformly bounded on $\mathbb{T}+i\nu$, we have
	$$
	||\hat{u}(\cdot,\lambda)||_{H(V^{*})}^{2}\leqslant\mathrm{const}||\hat{f}(\cdot,\lambda)||_{H(V^{*})}^{2}
	$$
	with $\mathrm{const}$ independent of $\lambda$.
	Integrating this inequality over $\lambda\in\mathbb{T}+i\nu$ and applying Parseval's identity \eqref{eq:2.3.9}, we arrive at the estimate \eqref{eq:2.3.12}. Thus, the function $u$ from \eqref{eq:2.3.11} belongs to $H_{\nu}(V)$ and solves the problem \eqref{eq:2.3.5}. It remains to check the uniqueness of the solution. Let $v\in H_{\nu}(V)$ be a solution of the problem \eqref{eq:2.3.5} with zero right-hand side. Then the function $\hat{v}(\cdot,\lambda)$ solves the equation \eqref{eq:2.3.1} with $\hat{f}\equiv0$ and hence equals zero for almost all $\lambda\in\mathbb{T}+i\nu$. Consequently, $v\equiv 0$.
\end{proof}

Let us find an asymptotics of the solution to problem \eqref{eq:2.3.5}. Let $\beta$ be such a positive number that in the rectangle $\{\lambda\in\mathbb{C}:0\leqslant \mathrm{Re}\lambda<2\pi, |\rIm\;\lambda|\leqslant\beta\}$ only the real poles of the resolvent $\lambda\mapsto R(\lambda,\mu)$ lie and formula \eqref{eq:ResWithout} holds. We number these poles by the index $k=1,\ldots,\varkappa$ with multiplicity taken into account. This means that the poles solving the equation $\mu_{j}(\lambda)=\mu$ simultaneously for several $j$ are repeated and numbered by distinct indices $k$. Set
\begin{equation}\label{eq:2.3.13}
	v_{k}(x^{*},n)=\exp(i\lambda_{k}n)\psi_{j}(x^{*},\lambda_{k}),
\end{equation}
where $j=j(k)$ is such an index that $\mu_{j}(\lambda_{k})=\mu$. From \eqref{eq_A_exp} it follows that $v_{k}$ satisfies the homogeneous equation \eqref{eq:2.3.5}. Let us introduce the cut-off function
\begin{equation}\label{eq:2.3.14}
	\eta(n)=\left\{\begin{array}{ll}
		1,&n\geqslant0,\\
		0,&n<0.
	\end{array}\right.
\end{equation}

\begin{thm}\label{th:2.3.8}
	Let $u\in H_{-\beta}(V)$ and $v\in H_{\beta}(V)$ be solutions of the equation \eqref{eq:2.3.5} with right-hand side $f\in H_{-\beta}(V)\cap H_{\beta}(V)$. Then the following identities hold
	\begin{equation}\label{eq:2.3.15}
		\begin{gathered}
			u=r+\eta\sum\limits_{k=1}^{\varkappa}c_{k}v_{k}, \qquad
			v=r-(1-\eta)\sum\limits_{k=1}^{\varkappa}c_{k}v_{k},
		\end{gathered}
	\end{equation}
	where the function $r$ belongs to $H_{-\beta}(V)\cap H_{\beta}(V)$. The constants $c_{k}=c_{k}(f)$ are computed by the formulas
	\begin{equation}\label{eq:2.3.16}
		c_{k}(f)=\frac{i(f,v_{k})_{H(V)}}{\mu_{j}'(\lambda_{k})}
	\end{equation}
	and are continuous functionals of $f\in H_{-\beta}(V)\cap H_{\beta}(V)$.
\end{thm}

\begin{proof}
We first assume that $f$ is a function with finite support. Then $\lambda\mapsto\hat{f}(\lambda)$ is a trigonometric polynomial, $u$ and $v$ are given by formulas of the form \eqref{eq:2.3.11} with $\nu=-\beta$ and $\nu=\beta$, respectively, and the integrand is a meromorphic $2\pi$-periodic function of $\lambda\in\mathbb{C}$. Therefore, taking into account the representation for the resolvent \eqref{eq:ResWithout}, the following formula holds
	\begin{equation}\label{eq:2.3.17}
		u(x^{*},n)=v(x^{*},n)+\frac{1}{\sqrt{2\pi}}\int\limits_{\Gamma}\exp(in\lambda)\sum\limits_{l=1}^{N}\frac{(\hat{f}(\lambda),\psi_{l}(\bar{\lambda}))_{H(V^{*})}\psi_{l}(\lambda)}{\mu_{l}(\lambda)-\mu}\mathrm{d}\lambda,
	\end{equation}
	where $\Gamma$ is a rectangular closed contour joining the points $-i\beta$, $2\pi-i\beta$, $2\pi+i\beta$, $i\beta$ and $-i\beta$. Here, without loss of generality, we assume that there are no poles of the resolvent on the contour $\Gamma$, i.e., the points $\lambda=0$ and $\lambda=2\pi$ are not poles. Otherwise, we shift all integration contours to the left by a small number, and by the periodicity of the resolvent $\lambda \mapsto R(\lambda, \mu)$ the integrals do not change. Hence, the integral to the right in \eqref{eq:2.3.17} can be computed by residues. At each pole, several terms of the sum over $l$ may contribute to the residue. The numbering introduced before formula \eqref{eq:2.3.13} takes this into account. Then
	$$
	u(x^{*},n)=v(x^{*},n)+i\sqrt{2\pi}\sum\limits_{k=1}^{\varkappa}\underset{\lambda=\lambda_{k}}{\Res}\exp(in\lambda)\frac{ \psi_{j}(\lambda)(\hat{f}(\lambda),\psi_{j}(\bar{\lambda}))_{H(V^{*})}}{\mu_{j}(\lambda)-\mu},
	$$
	where $j=j(k)$. Computing the residues, we obtain
	$$
	u=v+\sum\limits_{k=1}^{\varkappa}c_{k}v_{k}
	$$
	with coefficients
	$$
	c_{k}(f)=\frac{i\sqrt{2\pi}}{\mu'_{j}(\lambda_{k})}(\hat{f}(\lambda_{k}),\psi_{j}(\lambda_{k}))_{H(V^{*})}.
	$$
	Here the denominator does not vanish, since $\mu$ is assumed to be different from thresholds.
	Taking into account the expression \eqref{eq_fourier_V} for $\hat{f}$, we arrive at the formula \eqref{eq:2.3.16}. Moreover, the functional \eqref{eq:2.3.16} extends by continuity to the whole space $H_{-\beta}(V)\cap H_{\beta}(V)$. Set
	\begin{equation}\label{eq:2.3.18}
		r=u-\eta\sum\limits_{k=1}^{\varkappa}c_{k}v_{k}=v+(1-\eta)\sum\limits_{k=1}^{\varkappa}c_{k}v_{k}.
	\end{equation}
	Since $\eta v_{k}\in H_{-\beta}(V)$ and $(1-\eta)v_{k}\in H_{\beta}(V)$, the function $r$ belongs to $H_{-\beta}(V)\cap H_{\beta}(V)$. Thus, formulas \eqref{eq:2.3.15} are proved under the assumption that the support of $f$ is finite. Let us get rid of this requirement.
	
	Let $\{f^{(m)}\}$ be a sequence of functions with finite support, approximating the element $f$ in $H_{-\beta}(V)\cap H_{\beta}(V)$, and let $u^{(m)}\in H_{-\beta}(V)$, $v^{(m)}\in H_{\beta}(V)$ be solutions of the problem \eqref{eq:2.3.5} with the right-hand side $f^{(m)}$. Then formulas \eqref{eq:2.3.15} hold with $u$, $v$, and $r$ replaced by $u^{(m)}$, $v^{(m)}$, and $r^{(m)}$, respectively:
	\begin{equation}\label{eq:2.3.19}
		r^{(m)}=u^{(m)}-\eta\sum\limits_{k=1}^{\varkappa}c_{k}(f^{(m)})v_{k}=v^{(m)}+(1-\eta)\sum\limits_{k=1}^{\varkappa}c_{k}(f^{(m)})v_{k}.
	\end{equation}
	By the inequality \eqref{eq:2.3.12}, the sequence $\{u^{(m)}\}$ converges to $u$ in $H_{-\beta}(V)$, and $\{v^{(m)}\}$ converges to $v$ in $H_{\beta}(V)$. Hence, by the continuity of the functionals $c_{k}$ on $H_{-\beta}(V)\cap H_{\beta}(V)$ and formula \eqref{eq:2.3.19}, the sequence $\{r^{(m)}\}$ converges in $H_{-\beta}(V)\cap H_{\beta}(V)$; denote the limit by $r$. Passing to the limit in formula \eqref{eq:2.3.19}, we obtain \eqref{eq:2.3.18}, from which the statement of the theorem follows.
\end{proof}

\begin{rem}\label{rm:2.3.9}
	Formulas \eqref{eq:2.3.15} provide the asymptotics of the solution $u\in H_{-\beta}(V)$ as $n\rightarrow+\infty$ and that of the solution $v\in H_{\beta}(V)$ as $n\rightarrow-\infty$; the function $r$ plays the role of the remainder.
\end{rem}

\begin{rem}\label{rm:2.3.10}
	From \eqref{eq:2.3.15} the relation between the solutions $u$ and $v$ follows. Namely, if from the first formula \eqref{eq:2.3.15} we subtract the second one, we obtain
	\begin{equation}\label{eq:2.3.20}
		u(x^{*},n)=v(x^{*},n)+\sum\limits_{k=1}^{\varkappa}c_{k}(f)v_{k}(x^{*},n).
	\end{equation}
	This formula should be understood pointwise, since the functions involved belong to different classes. The sum on the right-hand side solves the homogeneous equation \eqref{eq:2.3.5}.
\end{rem}

\begin{rem}\label{rm:2.3.11}
	The results of Theorem {\rm\ref{th:2.3.8}} can be extended to arbitrary $\beta_{1}$ and $\beta_{2}$ such that $\beta_{1}<\beta_{2}$. For this sake one needs to know the expansion of the resolvent in a neighborhood of the poles $\lambda_{m}$, $m=1,\ldots,M$, located in the rectangle between the segments $\mathbb{T}+i\beta_{1}$ and $\mathbb{T}+i\beta_{2}$. In a neighborhood of such pole $\lambda_{m}$ the following representation holds (see, for example, {\rm\cite{KozMaz}}, Section A)
	$$
	R(\lambda,\mu)=\sum\limits_{j=1}^{J_{m}}\sum\limits_{p=1}^{\varkappa_{j}}(\lambda-\lambda_{m})^{-p}\sum\limits_{q=0}^{\varkappa_{j}-p}(\cdot,z_m^{(q,j))})_{H(V^{*})}\varphi_m^{(\varkappa_{j}-p-q,j)}+\mathfrak{H}(\lambda),
	$$
	where $\mathfrak{H}(\lambda)$ is the holomorphic part of the resolvent, $\{\varphi_m^{(q,j)}\}_{q=0}^{\varkappa_{j}-1}\subset H(V^{*}), j=1, \dots J_m,$ are the Jordan chains of the operator-function $\lambda\mapsto A(\lambda)-\mu I$ corresponding to the eigenvalue $\lambda_{m}$, and $z_m^{(q,j)}$ are certain elements of $H(V^{*})$. Repeating the arguments in the proof of Theorem {\rm\ref{th:2.3.8}}, after shifting the contour and computing residues, we obtain
	$$
	u(x^{*},n)=v(x^{*},n)+\sum\limits_{m=1}^{M}\sum\limits_{j=1}^{J_{m}}\sum\limits_{p=0}^{\varkappa_{j}-1}c_{m,p,j}v_{m,p,j}(x^{*},n),
	$$
	where $c_{m,p,j}$ are continuous functionals on $H_{\beta_{1}}(V)\cap H_{\beta_{2}}(V)$ and
	$$
	v_{m,p,j}(x^{*},n)=\exp(i\lambda_{m}n)\sum\limits_{q=0}^{p}\frac{(in)^{q}}{q!}\varphi^{(p-q,j)}(x^{*}).
	$$
\end{rem}

\begin{rem}\label{rm:2.3.12}
	By Proposition {\rm\ref{lm:2.3.2}}, in the vertical strip $\mathrm{Re}\;\lambda\in[0,2\pi)$ the resolvent $\lambda\mapsto R(\lambda,\mu)$ has finitely many poles; then, starting from some $\gamma>0$, the set $|\rIm\;\lambda|>\gamma$ contains no poles of the resolvent. From Remark {\rm\ref{rm:2.3.11}} it follows that for $\beta_{2}>\beta_{1}>\gamma$ or $\beta_{1}<\beta_{2}<-\gamma$ the solution $u\in H_{\beta}(V)$ of the problem with right-hand side $f\in H_{\beta_{1}}(V)\cap H_{\beta_{2}}(V)$ does not depend on $\beta\in [\beta_{1},\beta_{2}]$ .
\end{rem}
\subsection{Incoming and outgoing waves}\label{ss_Cyl_waves}
Let $\mu$ be a real spectral parameter coinciding with neither a threshold nor a flat band and let $\beta$ be the same number as in the previous section. Let also $\eta$ be the cut-off function in \eqref{eq:2.3.14}. We denote by $W$ the linear span of functions of the form $\eta v_{k}$, where $v_{k}$ are the functions \eqref{eq:2.3.13}, $k=1,\ldots,\varkappa$, and call it the space of waves in semi-cylinder. In this section we choose in $W$ a basis of waves, either ``incoming from $+\infty$'' or ``outgoing to $+\infty$''. For $t\in\mathbb{Z}$ set
$$
V(t)=\{(x^{*},n)\in V: n<t\}.
$$
For arbitrary functions $u,v:V \to \bC$ let us introduce the sesquilinear form
\begin{align}\label{eq:2.3.21}
	q_{t}(\eta u,\eta v)=(A(\eta u),\eta v)_{H(V(t))}-(\eta u, A(\eta v))_{H(V(t))}.
\end{align}
The following Proposition \ref{pr:2.3.14} implies, in particular, that this form is well defined, since  the summation in the inner products is carried out over a finite set 
$\{(x^*,n)\in V: -t_0\leqslant n<t\}$ with $t_{0}:=\max\limits_{e^*\in E^*}|\tau(e^*)|>0$.
\begin{prop}\label{pr:2.3.14}$\,$\\
	Let $u:V\to\bC$ be an arbitrary function, and let $\eta$ be the cut-off function in \eqref{eq:2.3.14}. Then 
	\begin{align*}
		\big(A(\eta u)\big)(x^*,n)=0 \text{ for } n<-t_0, \qquad  \big(A(\eta u)\big)(x^*,n)=\big(A u\big)(x^*,n) \text{ for } n \geqslant t_0.
	\end{align*}
\end{prop}
\begin{proof}
	By formula \eqref{eq_A_explicit_Cyl} we have
	\begin{align}\label{eq_A_explicit_Cyl2}
		\big(A(\eta u)\big)(x^*,n)=\sum\limits_{e^*\in E^*(x^*)}a(e^*)\Big(\big(\eta u\big)\big(x^*,n\big)-\big(\eta u\big)\big(t(e^*), n+\tau(e^*)\big)\Big) 
		+q(x^*)\big(\eta u\big)(x^*,n).
	\end{align}
	First let $n<-t_0$. Then $n<0$ and $n+\tau(e^*)<0$ for any edge $e^*\in E^*$. Therefore $\eta(n)=\eta(n+\tau(e^*))=0$ and all terms to the right in \eqref{eq_A_explicit_Cyl2} vanish.
	Now let $n\geqslant t_0$. Then $n>0$ and $n+\tau(e^*)\geqslant 0$ for any edge $e^*\in E^*$. Therefore $\eta(n)=\eta(n+\tau(e^*))=1$ and the expression in the right-hand side of \eqref{eq_A_explicit_Cyl2} coincides with $\big(Au\big)(x^*,n)$.
\end{proof}

Let $\mathbb{U}$ be the neighborhood of the real axis from formula \eqref{eq:2.3.2}. For each $\lambda\in\mathbb{U}$ let us introduce the function $u_{j}(\lambda):V\to \bC$ by the formula
$$
u_{j}(x^{*},n,\lambda)=\exp(in\lambda)\psi_{j}(x^{*},\lambda),
$$
where $\psi_j$ are the functions in \eqref{eq_A_lam_eigenfun}. From formula \eqref{eq_A_exp} the identity $Au_j(\lambda)=\mu_j(\lambda)u_j(\lambda)$ follows.
To compute the form \eqref{eq:2.3.21} on the waves $\eta v_{k}$, we need the following statement.

\begin{prop}\label{pr:2.3.15}
	For any $z_{1},z_{2}\in\mathbb{U}$ and $t\geqslant t_0$ the following identities hold
	\begin{equation}\label{eq:2.3.23}
		q_{t}(\eta u_{j}(z_{1}),\eta u_{l}(\bar{z}_{2}))=\exp(i(z_{1}-z_{2})t)(\psi_{j}(z_{1}),\psi_{l}(\bar{z}_{2}))_{H(V^{*})}\frac{\mu_{l}(z_{2})-\mu_{j}(z_{1})}{1-\exp(i(z_{1}-z_{2}))}.
	\end{equation}
	The left- and right-hand sides depend analytically on the pair $(z_{1},z_{2})\in \bU\times \bU$.
\end{prop}

\begin{proof}
	First assume that $\rIm(z_{1}-z_{2})>0$. Then
	$$
	\begin{gathered}
		(A(\eta u_{j}(z_{1})),\eta u_{l}(\bar{z}_{2}))_{H(V(t))}=(A(\eta u_{j}(z_{1})),\eta u_{l}(\bar{z}_{2}))_{H(V)}-\\
		-\sum\limits_{n=t}^{\infty}(A(\eta u_{j}(z_{1}))(\cdot,n),(\eta u_{l}(\bar{z}_{2}))(\cdot,n))_{H(V^{*})}.
	\end{gathered}
	$$
	Both terms on the right-hand side converge by the assumption $\rIm(z_{1}-z_{2})>0$. Taking Proposition \ref{pr:1.2.4} into account, in the first term on the right-hand side we transfer the operator $A$ to the second factor and again isolate the inner product in $H(V(t))$:
	$$
	\begin{gathered}
		(A(\eta u_{j}(z_{1})),\eta u_{l}(\bar{z}_{2}))_{H(V(t))}=(\eta u_{j}(z_{1}),A(\eta u_{l}(\bar{z}_{2})))_{H(V(t))}+\\
		+\sum\limits_{n=t}^{\infty}((\eta u_{j}(z_{1}))(\cdot,n),A(\eta u_{l}(\bar{z}_{2}))(\cdot,n))_{H(V^{*})}-\\
		-\sum\limits_{n=t}^{\infty}(A(\eta u_{j}(z_{1}))(\cdot,n),(\eta u_{l}(\bar{z}_{2}))(\cdot,n))_{H(V^{*})}.
	\end{gathered}
	$$
	Thus, the following identity holds
	\begin{align*}
		q_{t}(\eta u_{j}(z_{1}),\eta u_{l}(\bar{z}_{2}))=&\sum\limits_{n=t}^{\infty}((\eta u_{j}(z_{1}))(\cdot,n),(A\eta u_{l}(\bar{z}_{2}))(\cdot,n))_{H(V^{*})}-\\
		-&\sum\limits_{n=t}^{\infty}(A(\eta u_{j}(z_{1}))(\cdot,n),(\eta u_{l}(\bar{z}_{2}))(\cdot,n))_{H(V^{*})}.
	\end{align*}
	In both sums on the right-hand side the inequality $n\geqslant t\geqslant t_{0}$ holds, and by Proposition \ref{pr:2.3.14} we have $\big(A(\eta u)\big)(x^*,n)=\big(Au\big)(x^*,n)$, that is, $\eta$ can be omitted. Since the functions $u_{j}(\lambda)$ solve the homogeneous problem \eqref{eq:2.3.5} with $\mu=\mu_{j}(\lambda)$, we obtain
	$$
	\begin{gathered}
		q_{t}(\eta u_{j}(z_{1}),\eta u_{l}(\bar{z}_{2}))=\sum\limits_{n=t}^{\infty}(u_{j}(\cdot,n,z_{1}),\mu_{l}(\bar{z}_{2})u_{l}(\cdot,n,\bar{z}_{2}))_{H(V^{*})}-\nonumber\\
		-\sum\limits_{n=t}^{\infty}(\mu_{j}(z_{1})u_{j}(\cdot,n,z_{1}),u_{l}(\cdot,n,\bar{z}_{2}))_{H(V^{*})}=\\
		=\big(\overline{\mu_{l}(\bar{z}_{2})}-\mu_{j}(z_{1})\big)\sum\limits_{n=t}^{\infty}(u_{j}(\cdot,n,z_{1}),u_{l}(\cdot,n,\bar{z}_{2}))_{H(V^{*})}=\\
		=\big(\overline{\mu_{l}(\bar{z}_{2})}-\mu_{j}(z_{1})\big)(\psi_{j}(z_{1}),\psi_{l}(\bar{z}_{2}))_{H(V^{*})}\sum\limits_{n=t}^{\infty}\exp(i(z_{1}-z_{2})n)=\\
		=\big(\mu_{l}(z_{2})-\mu_{j}(z_{1})\big)(\psi_{j}(z_{1}),\psi_{l}(\bar{z}_{2}))_{H(V^{*})}\frac{\exp(i(z_{1}-z_{2})t)}{1-\exp(i(z_{1}-z_{2}))},
	\end{gathered}
	$$
	where the identity $\overline{\mu_{l}(\bar{z}_{2})}=\mu_{l}(z_{2})$ from \eqref{eq:2.3.2} has been used.
	
	Let us get rid of the requirement $\rIm(z_{1}-z_{2})>0$. In the left-hand side of \eqref{eq:2.3.23}, the summation is carried out over a finite set (see \eqref{eq:2.3.21}), hence it is a finite sum of analytic functions of $(z_{1},z_{2})\in\mathbb{U}\times\mathbb{U}$ and is itself analytic. We shall show that the right-hand side has the same property. Then, by uniqueness of analytic continuation, formula \eqref{eq:2.3.23} holds for all $(z_{1},z_{2})\in\mathbb{U}\times\mathbb{U}$.
	By Hartogs's theorem (see \cite{Her}), it is enough to check analyticity in each variable with the other variable fixed. Fix a point $z_{2}\in\mathbb{U}$. Clearly, the right-hand side of formula \eqref{eq:2.3.23} is analytic in $z_{1}$ in the domain $\bU$ except for the zeros $z_{1}= z_{2}+2\pi k$, $k \in \bZ$, of the function $1-\exp(i(z_1-z_2))$. By the analyticity of the left-hand side of \eqref{eq:2.3.23}, its right-hand side has removable singularities at these points. 
	This can be checked directly; let us restrict ourselves to the case $z_1=z_2$. By \eqref{eq:2.3.2} we have $$(\mu_{l}(z_{2})-\mu_{j}(z_{2}))(\psi_{j}(z_{2}),\psi_{l}(\bar{z}_{2}))_{H(V^{*})}=(\mu_{l}(z_{2})-\mu_{j}(z_{2}))\delta_{jl}=0.$$
	As $z_1\to z_2$, the following formulas hold
	\begin{align}\label{eq:2.3.24}
		(\mu_{l}(z_{2})-\mu_{j}(z_{1}))(\psi_{j}(z_{1}),\psi_{l}(\bar{z}_{2}))_{H(V^{*})}&=(c_{1}(z_{2})-\delta_{jl}\mu'_{j}(z_{2}))(z_{1}-z_{2})+O((z_{1}-z_{2})^{2}),\nonumber\\
		1-\exp(i(z_1-z_2))&=-i(z_{1}-z_{2})+O((z_{1}-z_{2})^{2}),
	\end{align}
	where $c_{1}(z_{2})=(\mu_{l}(z_{2})-\mu_{j}(z_{2}))(\psi'_{j}(z_{2}),\psi_{l}(\bar{z}_{2}))_{H(V^{*})}$. The zeros in the numerator and denominator of \eqref{eq:2.3.23} cancel, and the identity
	\begin{align}\label{eq_waves_biort}
		q_{t}(\eta u_{j}(z_{2}),\eta u_{l}(\bar{z}_{2}))=i(c_{1}(z_{2})-\delta_{jl}\mu'_{j}(z_{2}))
	\end{align}
	holds.
\end{proof}

In the space $W$ we introduce the basis of waves:
\begin{equation}\label{eq:2.3.25}
	w_{k}(x^{*},n):=\eta(n)\frac{u_{j(k)}(x^{*},n,\lambda_{k})}{\sqrt{|\mu_{j(k)}'(\lambda_{k})|}}=\eta(n)\frac{\exp(in\lambda_{k})\psi_{j(k)}(x^{*},\lambda_{k})}{\sqrt{|\mu_{j(k)}'(\lambda_{k})|}},\quad k=1,\ldots,\varkappa,
\end{equation}
where $\lambda_k$ are the real roots of the equations $\mu_{j}(\lambda)=\mu$, $j=1, \dots, |V^*|,$ from the interval $[0, 2\pi)$, numbered with multiplicity taken into account, and $j(k)$ is such an index that $\mu_{j(k)}(\lambda_k)=\mu$ (see the explanation to formula \eqref{eq:2.3.13}).

\begin{prop}\label{pr:2.3.16}$\,$\\
	For the waves \eqref{eq:2.3.25} the orthogonality and normalization conditions
	\begin{equation}\label{eq:2.3.26}
		iq_{t}(w_{k},w_{l})=\sigma_{k}\delta_{kl},
	\end{equation}
	hold, where $\sigma_{k}:=\sgn\mu'_{j(k)}(\lambda_{k})$.
\end{prop}

\begin{proof}
	By \eqref{eq:2.3.25} we write
	$$
	iq_{t}(w_{k},w_{l})=\frac{iq_{t}(\eta u_{j(k)}(\lambda_{k}),\eta u_{j(l)}(\lambda_{l}))}{\sqrt{|\mu_{j(k)}'(\lambda_{k})|\cdot|\mu_{j(l)}'(\lambda_{l})|}}
	$$
	and compute the right-hand side using \eqref{eq:2.3.23}.
	Since $\mu_{j(k)}(\lambda_{k})=\mu=\mu_{j(l)}(\lambda_{l})$, the numerator in the right-hand side of the corresponding formula \eqref{eq:2.3.23} vanishes. Since $\lambda_k, \lambda_l \in [0, 2\pi)$, if $\lambda_{k}\neq\lambda_{l}$ the denominator is nonzero and $iq_{t}(w_{k},w_{l})=0$.
	Consider the case $\lambda_{k}=\lambda_{l}$. Taking into account the numbering of the numbers $\lambda_k$, we have either $k=l$ or $j(k)\neq j(l)$. By \eqref{eq_waves_biort} 
	\begin{align}
		q_{t}(\eta u_{j(k)}(\lambda_{k}),\eta u_{j(l)}(\bar{\lambda}_{k}))=i(c_{1}(\lambda_{k})-\delta_{j(k),j(l)}\mu'_{j(k)}(\lambda_{k})).
	\end{align}
	Here $c_{1}(\lambda_{k})=(\mu_{j(l)}(\lambda_{k})-\mu_{j(k)}(\lambda_{k}))(\psi'_{j(k)}(\lambda_{k}),\psi_{j(l)}(\bar{\lambda}_{k}))_{H(V^{*})}=0$. Then, for $k\neq l$ we get zero on the right-hand side, and for $k=l$ we get $-i\mu'_{j(k)}(\lambda_{k})$. Combining the obtained formulas, we arrive at \eqref{eq:2.3.26}.
\end{proof}

In Proposition \ref{pr:2.3.16} we proved that the value $q_{t}(u,v)$ of the form \eqref{eq:2.3.21} on the waves $u,\;v$ does not depend either on the number $t$ or on the cut-off function $\eta$; hence, below we shall omit the index $t$. The form $iq(\cdot,\cdot)$ is Hermitian, that is, for any waves $u,\,v\in W$ the identity
$$
iq(u,v)=\overline{iq(v,u)}
$$
holds and, in particular, the value $iq(u,u)$ is real. We call waves $u$ and $v$ orthogonal if $q(u,v)=0$. A wave $u$ is called incoming if $iq(u,u)<0$, and outgoing if $iq(u,u)>0$. Thus, the wave $w_{k}$ in \eqref{eq:2.3.26} is incoming if $\mu'_{j(k)}(\lambda_{k})<0$, and outgoing if  $\mu'_{j(k)}(\lambda_{k})>0$.
\subsection{Problem in the cylinder with Dirichlet boundary conditions}\label{ss_Cyl_Dir}
The results of Sections \ref{ss_Fund_spec}-\ref{ss_Cyl_waves} are extended to the case of the problem \eqref{eq:2.2.1} with Dirichlet boundary conditions. In this section we describe the necessary changes in the arguments.
In this case the problem \eqref{eq:2.2.16} takes the form 
\begin{equation}\label{eq_Pencil_Dir}
	\left\{\begin{array}{ll}
		(A(\lambda)\hat{u})(x^{*},\lambda)-\mu \hat{u}(x^{*},\lambda)=\hat{f}(x^{*},\lambda),&x^{*}\in\mathring{V}^{*}\\
		\hat{u}(x^{*},\lambda)=\hat{g}(x^{*},\lambda),&x^{*}\in\partial V^{*}.
	\end{array}\right.
\end{equation}
We rewrite the problem \eqref{eq_Pencil_Dir} as an equation of the form $A^D(\lambda)w-\mu w=F$ for some operator $A^D(\lambda)$. Let $P:H(V^*)\to H(\mathring{V}^*)$ be the restriction operator to $\mathring{V}^*$ of a function defined on $V^*$. The operator $P^*:H(\mathring{V}^*)\to H(V^*)$ acts as extension by zero on $\partial V^*$ of a function defined on $\mathring{V}^*$. Then $PP^*=I$ in $H(\mathring{V}^*)$, and $P^* P$ is the orthogonal projection in $H(V^*)$ onto the subspace $\{v \in H(V^*): v(x^*)=0 \text{ for } x^* \in \partial V^*\}$. Similarly, we define the operator $Q:H(V^*)\to H(\partial V^*)$ of restriction to $\partial V^*$ of a function defined on $V^*$. For this operator all that was said for $P$ holds with $\mathring{V}$ replaced by $\partial V$, and $Q^*Q+P^*P=I$ on $H(V^*)$. In what follows, we will use analogous notations for any graph $G=(V,E)$ with boundary $\partial V$.
\begin{prop}\label{pr:Pencil_Dir}
	{\rm1.} The function $\hat{u} \in H(V^*)$ satisfies the system \eqref{eq_Pencil_Dir} if and only if $Q\hat{u}=\hat{g}$ and the function $w:=P\hat{u} \in H(\mathring{V}^*)$ satisfies the identity 
	$A^D(\lambda)w-\mu w=F$
	with the operator 
	\begin{align}\label{eq:ADlam}
		A^D(\lambda)=P A(\lambda)P^*
	\end{align}
	and the right-hand side $F=\hat{f}-P A(\lambda)Q^* \hat{g}$.\\
	{\rm2.} If $\mu$ is not an eigenvalue of the operator \eqref{eq:ADlam}, then problem \eqref{eq_Pencil_Dir} is uniquely solvable for any right-hand side, and its solution is given by the formula
	\begin{align}\label{eq:PencilDirSol}
		\hat{u}=P^* R^D(\lambda,\mu)F+Q^*\hat{g},
	\end{align}
	where $R^D(\lambda,\mu):=(A^D(\lambda)-\mu)^{-1}$.
\end{prop}
\begin{proof}
	We write the system \eqref{eq_Pencil_Dir} in the form
	\begin{align}\label{eq_Pencil_Dir2}
		P A(\lambda)\hat{u}-\mu P\hat{u}=\hat{f} , \quad  Q \hat{u}=\hat{g}.
	\end{align}
	Taking into account the identity $Q^*Q+P^*P=I$, this is equivalent to the system
	\begin{align}\label{eq_Pencil_Dir3}
		(P A(\lambda)P^* -\mu) P\hat{u}+P A(\lambda)Q^*Q\hat{u}=\hat{f} , \quad  Q \hat{u}=\hat{g}.
	\end{align}
	Using the second identity, we obtain that the function $w:=P\hat{u}$ satisfies the equation $P A(\lambda)P^* w -\mu w=F$ with the right-hand side $F=\hat{f}-P A(\lambda)Q^* \hat{g}$, which was to be proved. If $\mu$ is not an eigenvalue of the operator $A^D(\lambda)$, then $w=R^D(\lambda, \mu)F$. The solution of the original system is recovered by the formula
	$$
	\hat{u}=P^*P\hat{u}+Q^*Q\hat{u}=P^*R^D(\lambda, \mu)F+Q^*\hat{g}.
	$$
\end{proof}

The operator $A^D(\lambda):H(\mathring{V}^*)\to H(\mathring{V}^*)$ is a holomorphic $2\pi$-periodic function of the parameter $\lambda \in \bC$, and the identity $A^D(\bar{\lambda})=A^D(\lambda)^*$ holds. 
Thus, $A^D(\lambda)$ have the same properties as those of $A(\lambda)$ in Section \ref{ss_Fund_spec}. In particular, there exist functions $\{\lambda\mapsto\mu_{j}(\lambda)\in \bC\}_{j=1}^{N}$ and $\{\lambda\mapsto\psi_{j}(\lambda)\in H(V^*)\}_{j=1}^{N}$, $N=|\mathring{V}^{*}|$, analytic in a symmetric neighborhood $\mathbb{U}$ of the real axis, such that
\begin{align}\label{eq_A_lam_eigenfun_Dir}
	A^D(\lambda)\psi_{j}(\lambda)=\mu_{j}(\lambda)\psi_{j}(\lambda),\quad\forall \lambda\in\mathbb{U},
\end{align}
the vectors $\psi_{1}(\lambda),\ldots,\psi_{N}(\lambda)$ form a basis in $H(\mathring{V}^{*})$ and the relations \eqref{eq:2.3.2} hold. As this was in Section \ref{ss_Fund_spec}, the functions $\mu_{j}(\lambda)$ play the role of band functions. In what follows, we assume that $\mu$ coincides with neither thresholds nor flat bands, which are defined by analogy with Section \ref{ss_Fund_spec}.
\begin{prop}
	For $\lambda \in \bU$, the solution $\hat{u}$ of the problem \eqref{eq_Pencil_Dir} admits the representation
	\begin{align}\label{eq:ResWithout_Dir1}
		\hat{u}(x^{*},\lambda)=(Q^*\hat{g})(x^*, \lambda)+\sum\limits_{j=1}^{N}\frac{c_j(\lambda)(P^*\psi_{j})(x^*,\lambda)}{\mu_{j}(\lambda)-\mu}
	\end{align}
	with coefficients $c_j(\lambda):=(\hat{f}(\cdot, \lambda),\psi_{j}(\cdot,\bar{\lambda}))_{H(\mathring{V}^{*})}-(\hat{g}(\cdot, \lambda), B^N(\bar{\lambda})P^* \psi_j(\cdot, \bar{\lambda}))_{H(\partial V^*)}$, where $B^N(\lambda)$ is the operator from \eqref{eq:2.2.18}.
\end{prop}
\begin{proof}
	For $\lambda \in \bU$ the representation
	\begin{equation}\label{eq:ResWithout_Dir}
		R^D(\lambda,\mu)=\sum\limits_{j=1}^{N}\frac{(\cdot,\psi_{j}(\bar{\lambda}))_{H(\mathring{V}^{*})}\psi_{j}(\lambda)}{\mu_{j}(\lambda)-\mu}
	\end{equation}
	holds. Hence by \eqref{eq:PencilDirSol}
	$$
	\hat{u}(x^{*},\lambda)=(Q^*\hat{g})(x^*, \lambda)+\sum\limits_{j=1}^{N}\frac{(F(\cdot,\lambda),\psi_{j}(\cdot,\bar{\lambda}))_{H(\mathring{V}^{*})}(P^*\psi_{j})(x^*,\lambda)}{\mu_{j}(\lambda)-\mu}.
	$$
	It remains only to compute $(F(\cdot,\lambda),\psi_{j}(\cdot,\bar{\lambda}))_{H(\mathring{V}^{*})}$. Omitting the arguments of the functions for brevity and taking into account the expression for $F$, we have
	\begin{align}
		(F,\psi_{j})_{H(\mathring{V}^{*})}=(\hat{f},\psi_{j})_{H(\mathring{V}^{*})}-(P A(\lambda)Q^*\hat{g},\psi_{j})_{H(\mathring{V}^{*})}.
	\end{align}
	Next, taking into account the identity $A(\lambda)^*=A(\bar{\lambda})$ and formula \eqref{eq:2.2.18}, we obtain
	\begin{align}
		(P A(\lambda)Q^*\hat{g},\psi_{j})_{H(\mathring{V}^{*})}=(\hat{g},QA(\bar{\lambda})P^*\psi_{j})_{H(\partial V^*)}=(\hat{g}, B^N(\bar{\lambda})P^* \psi_j)_{H(\partial V^*)}.
	\end{align}
\end{proof}

We now turn to the discussion of the problem \eqref{eq:2.2.1} in cylinder with Dirichlet boundary conditions on $\partial V$:
\begin{equation}\label{eq:2.2.1_Dir}
	\left\{\begin{array}{ll}
		(Au)(x)-\mu u(x)=f(x),&x\in\mathring{V},\\
		u(x)=g(x),&x\in\partial V.
	\end{array}\right.
\end{equation}
Reasoning in the same way as in Section \ref{ss_Cyl_asym} and using Proposition \ref{pr:Pencil_Dir}, we obtain analogues of Theorems \ref{th:2.3.7} and \ref{th:2.3.8}. Below, by $H_{\beta}(\mathring{V})$ and $H_{\beta}(\partial V)$ we denote analogues of the space $H_{\beta}(V)$ with norms of the form \eqref{eq:2.3.7}.
\begin{thm}\label{th:2.3.7_Dir}
	Let the number $\beta\in\mathbb{R}$ be such that $\mathbb{T}+i\beta$ contains no poles of the operator-valued function $\lambda\mapsto R^D(\lambda,\mu)$. Then for any right-hand side $f\in H_{\beta}(\mathring{V}), g \in H_{\beta}(\partial V)$ there exists a unique solution $u\in H_{\beta}(V)$ of the problem \eqref{eq:2.3.5}. Such a solution is given by the formula
	\begin{equation}\label{eq:2.3.11_Dir}
		u(x^{*},n)=\frac{1}{\sqrt{2\pi}}\int\limits_{\mathbb{T}+i\beta}\exp(in\lambda)\left(\left(P^*R^D(\lambda,\mu)F\right)(x^{*},\lambda)+\left(Q^* \hat{g}\right)(x^{*},\lambda)\right)\mathrm{d}\lambda,
	\end{equation}
	and the estimate
	\begin{equation}\label{eq:2.3.12_Dir}
		\|u;H_{\beta}(V)\|\leqslant\mathrm{const}\left(\|f;H_{\beta}(\mathring{V})\|+\|g;H_{\beta}(\partial V)\|\right)
	\end{equation}
	holds.
\end{thm}

An asymptotics of the solution to problem \eqref{eq:2.3.5} is derived by means of formula \eqref{eq:ResWithout_Dir1} and shifting the contour in the integral \eqref{eq:2.3.11_Dir}.  
Let, as before, $\beta$ be such a positive number that in the rectangle $\{\lambda\in\mathbb{C}:0\leqslant \mathrm{Re}\lambda<2\pi, |\rIm\;\lambda|\leqslant\beta\}$ is a subset of $\bU$ and contains only the real poles of the resolvent $\lambda\mapsto R^D(\lambda,\mu)$.
We number these poles by the index $k=1,\ldots,\varkappa$ with multiplicity taken into account. Set
\begin{equation}\label{eq:2.3.13_Dir}
	v_{k}(x^{*},n):=\exp(i\lambda_{k}n)P^*\psi_{j}(x^{*},\lambda_{k}),
\end{equation}
where $j=j(k)$ is such an index that $\mu_{j}(\lambda_{k})=\mu$. Let also $\eta$ be the function in \eqref{eq:2.3.14}.
\begin{thm}\label{th:2.3.8_Dir}
	Let $u\in H_{-\beta}(V)$ and $v\in H_{\beta}(V)$ be solutions to problem \eqref{eq_Pencil_Dir} with right-hand side $f\in H_{-\beta}(\mathring{V})\cap H_{\beta}(\mathring{V})$ and $g \in H_{-\beta}(\partial V)\cap H_{\beta}(\partial V)$. Then the identities 
	\begin{equation}\label{eq:2.3.15_Dir}
		\begin{gathered}
			u=r+\eta\sum\limits_{k=1}^{\varkappa}c_{k}v_{k}, \qquad
			v=r-(1-\eta)\sum\limits_{k=1}^{\varkappa}c_{k}v_{k},
		\end{gathered}
	\end{equation}
hold with the function $r$ belonging to $H_{-\beta}(V)\cap H_{\beta}(V)$. The constants $c_{k}=c_{k}(f,g)$ are computed by the formulas
	\begin{equation}\label{eq:2.3.16_Dir}
		c_{k}(f,g)=\frac{i(f,P v_{k})_{H(\mathring{V})}-i(g,B^N v_k)_{H(\partial V)}}{\mu_{j}'(\lambda_{k})}
	\end{equation}
	and are continuous functionals of $$\{f,g\}\in \left(H_{-\beta}(\mathring{V})\times H_{-\beta}(\partial V)\right)\cap \left(H_{\beta}(\mathring{V})\times H_{\beta}(\partial V)\right).$$
\end{thm}

Let us denote by $W$ the linear span of functions of the form $\eta v_k$, where $v_k$ are the functions from \eqref{eq:2.3.13_Dir}, and call it the space of waves in semi-cylinder. We now define the waves incoming from $+\infty$ and outgoing to $+\infty$. To this end, for arbitrary functions $u,v:V \to \bC$ let us introduce the sesquilinear form
\begin{align}\label{eq:2.3.21_Dir}
	q_{t}(\eta u,\eta v)=&(A(\eta u),\eta v)_{H(\mathring{V}(t))}+(\eta u, -B^N (\eta v))_{H(\partial V(t))}\\-&(\eta u, A(\eta v))_{H(\mathring{V}(t))}-(-B^N(\eta u),  \eta v)_{H(\partial V(t))},
\end{align}
where $\mathring{V}(t):=\mathring{V}\cap V(t)$ and $\partial V(t)=\partial V\cap V(t)$. 

For each $\lambda\in\mathbb{U}$ let the function $u_{j}(\lambda):V\to \bC$ be acting as
$$
u_{j}(x^{*},n,\lambda)=\exp(in\lambda)P^*\psi_{j}(x^{*},\lambda),
$$
where $\psi_j$ are the functions in \eqref{eq_A_lam_eigenfun_Dir}. 
The functions $u_{j}(\lambda)$ vanish on $\partial V$. From formula \eqref{eq_A_exp} and Proposition \ref{pr:Pencil_Dir} the identity $Au_j(\lambda)=\mu_j(\lambda)u_j(\lambda)$ holds on $\mathring{V}$. Taking these remarks into account, one can repeat the proof of Proposition \ref{pr:2.3.15}.
For any $z_{1},z_{2}\in\mathbb{U}$ and $t\geqslant t_0$ the following identities hold:
\begin{equation}\label{eq:2.3.23_Dir}
	q_{t}(\eta u_{j}(z_{1}),\eta u_{l}(\bar{z}_{2}))=\exp(i(z_{1}-z_{2})t)(\psi_{j}(z_{1}),\psi_{l}(\bar{z}_{2}))_{H(\mathring{V}^{*})}\frac{\mu_{l}(z_{2})-\mu_{j}(z_{1})}{1-\exp(i(z_{1}-z_{2}))}.
\end{equation}
Finally, in the space $W$ we introduce a basis of waves:
\begin{equation}\label{eq:2.3.25_Dir}
	w_{k}(x^{*},n):=\eta(n)\frac{u_{j(k)}(x^{*},n,\lambda_{k})}{\sqrt{|\mu_{j(k)}'(\lambda_{k})|}}=\eta(n)\frac{\exp(in\lambda_{k})P^*\psi_{j(k)}(x^{*},\lambda_{k})}{\sqrt{|\mu_{j(k)}'(\lambda_{k})|}},\quad k=1,\ldots,\varkappa.
\end{equation}
Here $\lambda_k$ are the real roots of the equations $\mu_{j}(\lambda)=\mu$ with  $j=1, \dots, |\mathring{V}^*|$, belonging to the interval $[0, 2\pi)$ and numbered with multiplicity taken into account. Let $j(k)$ be such an index that $\mu_{j(k)}(\lambda_k)=\mu$. From formula \eqref{eq:2.3.23_Dir} it follows that the waves \eqref{eq:2.3.25_Dir} satisfy the orthogonality and normalization conditions
\begin{equation}\label{eq:2.3.26_Dir}
	iq_{t}(w_{k},w_{l})=\sigma_{k}\delta_{kl},
\end{equation}
where $\sigma_{k}:=\sgn\mu'_{j(k)}(\lambda_{k})$.

\section{Problem in a domain with outlets to infinity}\label{part:3}
Let us call a graph $G=(V,E)$ a \emph{waveguide} if it consists of a finite number of semi-cylinders $\Pi^{+}_{1},\ldots,\Pi^{+}_{\exnum}$, connected by a finite number of edges to a finite resonator $G_c$. Let us give a rigorous definition. We assume that a set of cylinders $\Pi_{1},\ldots,\Pi_{\exnum}$ is given; each cylinder $\Pi_r$ is determined by the set of vertices $V_{r}$, the set of edges $E_{r}$, and the period $\cT_{r}$. This allows us to introduce in the cylinder local coordinates  $(x_{r}^{*},n_{r})$ for vertices and $(e_{r}^{*},n_{r})$ for edges. Here $n_{r}$ is the number of the periodicity cell, $x_{r}^{*}\in V_{r}^{*}$ and $e_{r}^{*}\in E_{r}^{*}$ are, respectively, the vertices and edges of the fundamental graph $(V_{r}^{*},E_{r}^{*})$
(see formulas \eqref{eq_coord_V} and \eqref{eq_coord_E}). Then the semi-cylinder $\Pi^{+}_{r}$ is obtained from the cylinder $\Pi_{r}$ by removing the vertices with negative $n_{r}$ and the edges incident to them; the sets of vertices and edges of the semi-cylinder $\Pi^{+}_{r}$ will be denoted by $V^+_r$ and $E^+_r$, respectively. The resonator $G_c$ is assumed to be an arbitrary finite graph. Its vertices are connected by finitely many edges to the sets $\{(x_r^*,0), x_r* \in V_r^*\}\subset V_r^+$, $r=1, \dots, \exnum$.

On the graph $G$ we distinguish the boundary $\partial V \subset V$ (see the description after Proposition \ref{pr:1.2.4}). We also assume that in each cylinder $\Pi_r$ a periodic boundary $\partial V_r$ is distinguished (see the beginning of Section \ref{ss_Cyl}) and that, in each semi-cylinder $\Pi_r^+$, the boundary $\partial V$ of the waveguide $G$ coincides with the boundary $\partial V_r$ of the cylinder $\Pi_r$, i.e., $\partial V \cap V_r^+=\partial V_r\cap V_r^+$. Set $\mathring{V}:=V\setminus\partial V$.

We consider on the graph $G$ a problem of the form \eqref{eq:1.3.1}, but restrict ourselves to the case of Dirichlet boundary conditions:
\begin{align}\label{eq:3.1.1}
	\left\{\begin{array}{ll}
		(Au)(x)-\mu u(x)=f(x),&x\in \mathring V,\\
		u(x)=g(x),&x\in\partial V,
	\end{array}\right.
\end{align}
where $\mu \in \bR$ is the spectral parameter, $A$ is an operator of the form $-\mathrm{div}a\nabla+q$, and the functions $f$ and $g$ are given functions on $\mathring V$ and $\partial V$. The case $\partial V=\varnothing$ corresponds to the problem without boundary conditions; in this case the subsequent arguments can be naturally simplified. 
As before, it is assumed that $q$ is a real-valued function on the set of vertices (a potential) and that $a$ is a real-valued even function on the set of edges (a metric).

Assume that at infinity, in each cylindrical outlet $G~\cap~\Pi^{+}_{r},\;r=1,\ldots\exnum$, the functions $a(e)$ and $q(x)$ stabilize to functions $a_r(e_{r}^{*})$ and $q_r(x_{r}^{*})$ that do not depend on $n_r$. This means that for all $r=1,\ldots\exnum$, $e_{r}^{*}\in E_{r}^{*}$, and $x_{r}^{*}\in V_{r}^{*}$ the formulas
\begin{equation}\label{eq:3.1.4}
	\begin{gathered}
		|a(e_{r}^{*},n_{r})-a_r(e_{r}^{*})|\rightarrow0,\\
		|q(x_{r}^{*},n_{r})-q_r(x_{r}^{*})|\rightarrow0
	\end{gathered}
\end{equation}
hold as $n_{r}\rightarrow+\infty$.

The stabilization conditions \eqref{eq:3.1.4} imply that the functions $a(e)$ and $q(x)$ are bounded on $E$ and $V$, respectively. Therefore, by Proposition \ref{pr:1.2.4}, the operator
\begin{align}
	A: H(V)\to H(V),
\end{align}
acting by the formula $Au=-\mathrm{div}a\nabla u+q u$, is bounded and self-adjoint. The operator 
\begin{align}\label{eq:oper}
	A^D:=P A P^*: H(\mathring{V})\to H(\mathring{V})
\end{align}
corresponds to the problem with Dirichlet boundary conditions and has the same properties. Here, as before, $P$ stands for the restriction operator from $V$ to $\mathring{V}$, and $P^*$ stands for the extension by zero from $\mathring{V}$ to $V$.

For any $u,v\in H(V)$, the Green's formula
\begin{equation}\label{eq:3.1.3}
	(Au,v)_{H(\mathring{V})}+(u, -B^N v)_{H(\partial V)}=(u,Av)_{H(\mathring{V})}+(-B^N u,  v)_{H(\partial V)}
\end{equation}
holds, where $B^N$ is the boundary operator \eqref{eq:1.3.3}.

\subsection{Fredholm property of the problem}\label{subsec:3.1}
Let $A^{(0)}$ stand for the difference operator $A^{(0)}:= -\mathrm{div}a^{(0)}\nabla+q^{(0)}$, whose coefficients $a^{(0)}$ and $q^{(0)}$ in each cylindrical outlet $\Pi^{+}_{r}$ coincide with the limiting coefficients $a_r$ and $q_r$ from \eqref{eq:3.1.4}, i.e., $a^{(0)}(e)=a_r(e)$ for $e \in E_r^+$ and $q^{(0)}(x)=q_r(x)$ for $x \in V_r^+$, $r=1, \dots, \exnum$. On the remaining vertices and edges we assume, for definiteness, that the coefficients of the operators $A$ and $A^{(0)}$ coincide. The operator $A^{(0)}$ is called the principal part of $A$. 
In each cylinder $\Pi_{r}$ we consider the model problem
\begin{align}
	(A_r u)(x_{r}^{*},n_{r})-\mu u(x_{r}^{*},n_{r})&=f(x_{r}^{*},n_{r}),\quad x_{r}^{*}\in \mathring{V}_{r}^{*},\;n_{r}\in\mathbb{Z},\label{eq:3.1.5_Inn}\\
	u(x_{r}^{*},n_{r})&=g(x_{r}^{*},n_{r}),\quad x_{r}^{*}\in \partial V_{r}^{*},\;n_{r}\in\mathbb{Z},\label{eq:3.1.5_Bound}
\end{align}
where $A_r$ is the difference operator $A_r=-\mathrm{div}a_r\nabla+q_r$. This problem is studied by the methods of Section \ref{s_ProbCyl}. Let us associate with it the operator $A^D_r(\lambda)$, defined by a formula of the form \eqref{eq:ADlam}; see also \eqref{eq_A_explicit_lam}. The solvability of the problem \eqref{eq:3.1.5_Inn}, \eqref{eq:3.1.5_Bound} is established in Theorem \ref{th:2.3.7_Dir}.

For $\nu\in\mathbb{R}$ we introduce the space $H_{\nu}(V)$ of functions on $V$ with the norm
\begin{equation}\label{eq:3.1.6}
	\|u;H_{\nu}(V)\|:=\left(\sum\limits_{x\in V}\exp(2\nu|x|)|u(x)|^{2}\right)^{1/2},
\end{equation}
where $x \mapsto |x|$ is a function on $V$ that coincides on each semi-cylinder $V\cap V^{+}_{r}$ with the map $(x_{r}^{*},n_{r})\mapsto n_{r}$ and is equal to zero at the remaining finite number of vertices. For every $\nu$ the operator $u\mapsto \{A u-\mu u; B^D u\}$ of problem \eqref{eq:3.1.1} implements the continuous mapping
\begin{equation}\label{eq:3.1.7}
	\mathcal{L}_{\nu}(\mu):\;H_{\nu}(V)\longrightarrow H_{\nu}(\mathring{V})\times H_\nu(\partial V)=:\cH_\nu(\mathring{V},\partial V).
\end{equation}
An operator is called Fredholm if its range is closed and its kernel and cokernel (a direct complement to the range) are finite-dimensional. The next theorem gives a criterion of Fredholmness for the operator \eqref{eq:3.1.7}. 
Let $\mathcal{L}_{\nu}^{(0)}(\mu)$ stand for the operator \eqref{eq:3.1.7} obtained by replacing $A$ with $A^{(0)}$. As before, $\mathbb{T}$ stands for the interval $[0, 2\pi)$.

\begin{thm}\label{th:3.1.1}
	Let $\mu$ be distinct from all flat bands of the problems \eqref{eq:3.1.5_Inn}, \eqref{eq:3.1.5_Bound} for $r=1,\dots,d$. Then the mapping \eqref{eq:3.1.7} is Fredholm if and only if the interval $\mathbb{T}+i\nu:=\{\lambda \in \bC: \rRe \lambda \in [0, 2\pi), \, \rIm \lambda=\nu\}$ contains no poles of the resolvents $\lambda\mapsto R^D_{r}(\lambda,\mu):=(A^D_r(\lambda)-\mu)^{-1}$ for all $r=1,\ldots,\exnum$.
\end{thm}

We split the proof into several steps. First we obtain an a priori estimate for a solution $u\in H_{\nu}(V)$ to problem \eqref{eq:3.1.1} with $\{f,g\}\in \cH_\nu(\mathring{V},\partial V)$. Then, with the use of this estimate, we establish the closedness of the range $\rran\mathcal{L}_{\nu}(\mu)$ and verify the finite-dimensionality of $\ker\mathcal{L}_{\nu}(\mu)$. Finally, we construct a right regularizer for $\mathcal{L}_{\nu}(\mu)$, which is used for proving  $\coker\mathcal{L}_{\nu}(\mu)$ to be finite-dimensional.

\paragraph{1. A priori estimate of solutions.}
Assume that the functions $u \in H_{\nu}(V)$ and $\{f,g\}\in \cH_\nu(\mathring{V},\partial V)$ satisfy the equation \eqref{eq:3.1.1}, and that the interval $\mathbb{T}+i\nu$ contains no poles of the resolvents $\lambda\mapsto R^D_{r}(\lambda,\mu)$ for all $r=1,\ldots,\exnum$. The main result of this step is the estimate
\begin{equation}\label{eq:3.1.8}
	\|u;H_{\nu}(V)\|\leqslant c(\|f;H_{\nu}(\mathring{V})\|+\|g;H_{\nu}(\partial V)\|+\|u;H_{\nu}(W)\|),
\end{equation}
where $W$ is some finite subset of $V$. Here and below in the proof we denote by $c$ various constants, independent of $u, f$, and $f$. We derive the inequality \eqref{eq:3.1.8} from local estimates of the form \eqref{eq:2.3.12_Dir} in the cylindrical outlets of the waveguide. To this end, for each $r=1, \dots, \exnum$ we introduce on $V$ the cutoff function 
\begin{equation}\label{eq:3.1.10}
	\zeta_{r}(x)=\left\{\begin{array}{ll}
		1,& x=(x_{r}^{*},n_{r})\in V\cap V^{+}_{r},\,n_{r}\geqslant t,\\
		0,&\text{otherwise},
	\end{array}\right.
\end{equation}
where $t \in \bN$ is a sufficiently large number, to be fixed later. We also set $\zeta_{0}=1-\zeta_{1}-\ldots-\zeta_{\exnum}$ and represent the function $u$ as $u=\zeta_0 u +\zeta_1 u+\dots+\zeta_\exnum u$. The support of the function $\zeta_{r}u$, $r=1, \dots, \exnum$, is contained in $V\cap V^{+}_{r}$. Therefore one can introduce a function on $V_r$ that coincides with $\zeta_{r}u$ on $V_r^+$ and is zero on the remaining part of $V_r$. We also denote the resulting function by $\zeta_{r}u$ (the context precludes confusion) and say that the function $\zeta_{r}u$ is transferred from the waveguide $V$ into the cylinder $V_r$.
Now the norm \eqref{eq:3.1.6} is represented as
\begin{equation}\label{eq:3.1.9}
	\|u;H_{\nu}(V)\|=\left(\|\zeta_{0}u;H_{\nu}(V)\|^{2}+\sum\limits_{r=1}^{\exnum}\|\zeta_{r}u;H_{\nu}(V_r)\|^{2}\right)^{1/2}.
\end{equation}

We estimate separately each term to the right in \eqref{eq:3.1.9}. The cutoff $\zeta_{0}$ has finite support, and hence we obtain
\begin{equation}\label{eq:3.1.11}
	\|\zeta_{0}u;H_{\nu}(V)\|=\|u;H_{\nu}(W^{1})\|,
\end{equation}
where $W^{1}={\rm supp}(\zeta_{0})$ is a finite subset of $ V$. We now turn to estimating the function $\zeta_{r}u$, $r=1, \dots, \exnum$. We have
$$
(A-\mu)(\zeta_{r}u)=\zeta_{r}f+[A,\zeta_{r}]u,
$$
where $[\cdot,\cdot]$ is the commutator, $[A,\zeta_{r}]u:=A\zeta_{r}u-\zeta_{r}Au$. Set $\Delta A=A^{(0)}-A$, where $A^{(0)}$ is the principal part of the operator $A$. The system
\begin{align}
	(A^{(0)}-\mu)(\zeta_{r}u)&=F \text{ on } \mathring{V},\label{eq:3.1.12_Inn}\\
	\zeta_{r}u&=\zeta_{r}g \text{ on } \partial V,\label{eq:3.1.12_Bound}
\end{align}
holds with $F:=\zeta_{r}f+[A,\zeta_{r}]u+\Delta A(\zeta_{r}u)$.
The supports of the functions  $\zeta_r u$, $\zeta_{r}f$, and $\zeta_r g$ lie in $V\cap V_r^+$; by Proposition \ref{pr:2.3.14} this is also true for the functions $[A,\zeta_{r}]u$, $\Delta A(\zeta_{r}u)$ for sufficiently large $t$. Let us transfer the functions $\zeta_r u$, $F$, and $\zeta_r g$ from $V$ to $V_r$ and keep their previous notations. On  $V\cap V_r^+$ the coefficients of the operators $A^{(0)}$ and $A_r$ coincide; therefore the obtained functions satisfy the system \eqref{eq:3.1.5_Inn}, \eqref{eq:3.1.5_Bound} with $u$, $f$, and $g$ replaced by $\zeta_r u$, $F$, and $\zeta_r g$, respectively. By Theorem \ref{th:2.3.7_Dir} the inequality
\begin{equation}\label{eq:3.1.13}
	\|\zeta_{r}u; H_{\nu}(V_{r})\|\leqslant c\left(\|F; H_{\nu}(\mathring{V}_{r})\|+\|\zeta_r g; H_{\nu}(\partial V_{r})\|\right).
\end{equation}
holds.
We estimate the norm in $H_{\nu}(\mathring{V}_{r})$ of each term in the definition of $F$. Clearly,
\begin{equation}\label{eq:3.1.14}
	\|\zeta_{r}f;H_{\nu}(\mathring{V}_{r})\|\leqslant \|f;H_{\nu}(\mathring{V})\|.
\end{equation}
Next, by Proposition \ref{pr:2.3.14} the support of the function $[A,\zeta_{r}]u$ is contained in a finite set $W_{r}^{2}\subset V$, which is independent of $u$. Therefore, the commutator $[A,\zeta_{r}]$ is a finite-dimensional operator and, consequently,
\begin{equation}\label{eq:3.1.15}
	\|[A,\zeta_{r}]u; H_{\nu}(\mathring{V}_{r})\|\leqslant c\|u;H_{\nu}(W_{r}^{2})\|.
\end{equation}
Let us consider the last term in the definition of $F$. Set $t_{r}:=\max\limits_{e\in E_{r}}|\tau_{r}(e)|$, where $\tau_{r}$ is the edge index function for the cylinder $\Pi_{r}$. For a given $\varepsilon>0$ we choose the number $t$ in the definitions of $\zeta_{r}$ such that for $n_r\geqslant t-t_{r}$ the inequalities
$$
\max\limits_{e_{r}^{*}\in E_{r}^*}|a(e_{r}^{*},n_{r})-a_r(e_{r}^{*})|<\varepsilon,\qquad\max\limits_{x_{r}^{*}\in V_{r}^*}|q(x_{r}^{*},n_{r})-q_r(x_{r}^{*})|<\varepsilon
$$
hold. Then, taking the definition of $\Delta A$ into account, we obtain 
\begin{equation}\label{eq:3.1.16}
	\|\Delta A(\zeta_{r}u);H_{\nu}(\mathring{V}_{r})\|\leqslant c\varepsilon \|u; H_{\nu}(V)\|,
\end{equation}
where the constant $c$ depends neither on $u,f,g$ nor on $t$, and the quantity $\varepsilon$ can be made arbitrarily small by choosing $t$. From \eqref{eq:3.1.13} -- \eqref{eq:3.1.16} it follows that
\begin{equation}\label{eq:3.1.17}
	\|\zeta_{r}u;H_{\nu}(V_{r})\|\leqslant c(\|f;H_{\nu}(\mathring{V})\|+\|g;H_{\nu}(\partial V)\|+\|u;H_{\nu}(W_{r}^{2})\|)+c\varepsilon \|u;H_{\nu}(V)\|.
\end{equation}
By \eqref{eq:3.1.11} and \eqref{eq:3.1.17} for $r=1,\ldots,\exnum$ we have
\begin{equation}\label{eq:3.1.18}
	\begin{gathered}
		\|u;H_{\nu}(V)\|\leqslant \|\zeta_{0}u; H_{\nu}(V)\|+\sum\limits_{r=1}^{\exnum}\|\zeta_{r}u; H_{\nu}(V_{r})\|\leqslant \\
		\leqslant c(\|f;H_{\nu}(\mathring{V})\|+\|g;H_{\nu}(\partial V)\|+\|u;H_{\nu}(W^{1})\|+\sum\limits_{r=1}^{\exnum}\|u;H_{\nu}(W_{r}^{2})\|)+c\varepsilon\|u;H_{\nu}(V)\|.
	\end{gathered}
\end{equation}
Then
\begin{equation}\label{eq:3.1.19}
	\|u;H_{\nu}(V)\|\leqslant c(\|f;H_{\nu}(\mathring{V})\|+\|g;H_{\nu}(\partial V)\|+\|u;H_{\nu}(W)\|)+c\varepsilon\|u;H_{\nu}(V)\|,
\end{equation}
where
$$
W:=W^{1}\cup\bigcup\limits_{r=1}^{\exnum}W_{r}^{2}
$$
is a finite set. 
Let us choose $t$ so that $c\varepsilon<1$. Now \eqref{eq:3.1.8} follows from \eqref{eq:3.1.19}.
\paragraph{2. Closedness of the range and finite-dimensionality of the kernel of the operator \eqref{eq:3.1.7}.}
We begin by verifying the finite-dimensionality of the space $\ker\mathcal{L}_{\nu}(\mu)$. For an element $u\in\ker\mathcal{L}_{\nu}(\mu)$, the estimate \eqref{eq:3.1.8} takes the form
\begin{equation}\label{eq:3.1.20}
	\|u;H_{\nu}(V)\|\leqslant c\|u;H_{\nu}(W)\|.
\end{equation}
Since the set $W\subset V$ is finite, this estimate implies that the unit ball in the space $\ker\mathcal{L}_{\nu}(\mu)$ is compact. Only a finite-dimensional space has this property.

We now verify the closedness of the range $\rran\mathcal{L}_{\nu}(\mu)$ of the operator \eqref{eq:3.1.7}. Let $\mathcal{H}$ stand for a direct complement in $H_{\nu}(V)$ to the subspace $\ker \mathcal{L}_{\nu}(\mu)$. We show that for $u\in\mathcal{H}$ the inequality
\begin{equation}\label{eq:3.1.21}
	\|u;H_{\nu}(V)\|\leqslant c\left(\|(A-\mu)u;H_{\nu}(\mathring{V})\|+\|u;H_{\nu}(\partial V)\|\right)
\end{equation}
follows from \eqref{eq:3.1.8}. Suppose the contrary; then there exists a sequence $\{u_{m}\}\subset\mathcal{H}$ such that
\begin{align}
	\|u_{m};H_{\nu}(V)\|=1,\qquad\|(A-\mu)u_m;H_{\nu}(\mathring{V})\|+\|u_m;H_{\nu}(\partial V)\|\rightarrow0.
\end{align}
Since $W$ is finite and the sequence $\{u_{m}\}$ is bounded in the norm of $H_{\nu}(W)$, one can extract from $\{u_{m}\}$ a subsequence convergent in $H_{\nu}(W)$; for  simplicity of notations we assume that the sequence $\{u_{m}\}$ itself converges in $H_{\nu}(W)$. By \eqref{eq:3.1.8} we have
$$
\|u_{m}-u_{k};H_{\nu}(V)\|\leqslant c\big(\|(A-\mu)(u_{m}-u_{k});H_{\nu}(\mathring{V})\|+\|u_{m}-u_{k};H_{\nu}(\partial V)\|+\|u_{m}-u_{k};H_{\nu}(W)\|\big).
$$
Each term on the right-hand side tends to zero; hence the sequence $\{u_{m}\}$ is a Cauchy one in $H_{\nu}(V)$ and has a limit $u\in\mathcal{H}\subset H_{\nu}(V)$. Next, using the boundedness of the operator $A:H_{\nu}(V)\to H_{\nu}(\mathring{V})$, we obtain
\begin{align*}
	\|(A-\mu)u;H_{\nu}(\mathring{V})\|&=\lim\limits_{m\to \infty}\|(A-\mu)u_{m};H_{\nu}(\mathring{V})\|=0,\\
	\|u;H_{\nu}(\partial V)\|&=\lim\limits_{m\to \infty}\|u_m;H_{\nu}(\partial V)\|=0.
\end{align*}
Thus the element $u$ belongs to $\ker\mathcal{L}_{\nu}(\mu)\cap\mathcal{H}$ and is therefore zero. On the other hand, $\|u;H_{\nu}(V)\|=\lim\limits_{m\to \infty}\|u_{m};H_{\nu}(V)\|=1$. The contradiction proves \eqref{eq:3.1.21}. The inequality \eqref{eq:3.1.21} means that the mapping $\mathcal{L}_{\nu}(\mu):\mathcal{H}\rightarrow\rran\mathcal{L}_{\nu}(\mu)$ is an isomorphism; consequently, $\rran\mathcal{L}_{\nu}(\mu)$ is a closed subspace of $\cH_{\nu}(\mathring{V}, \partial V)$.

\paragraph{3. Right regularizer.}
On this step we construct a right regularizer $\cR_{\nu}(\mu)$ for the operator \eqref{eq:3.1.7}, that is, a bounded operator $\cR_{\nu}(\mu):\cH_{\nu}(\mathring{V}, \partial V)\rightarrow H_{\nu}(V)$ such that $\mathcal{L}_{\nu}(\mu)\cR_{\nu}(\mu)=I+\mathcal{K}$, where $I$ and $\mathcal{K}$ are the identity and a compact operator in $\cH_{\nu}(\mathring{V}, \partial V)$, respectively.

We first show that the operator $\Delta \cL:=\cL_\nu^{(0)}-\cL_\nu: H_{\nu}(V) \rightarrow \cH_{\nu}(\mathring{V}, \partial V)$ is compact for any $\nu\in\mathbb{R}$. Indeed, $\Delta \cL u=\{\Delta A u, 0\}$, where $\Delta A$ is the difference operator introduced before formula \eqref{eq:3.1.12_Inn}. Using the partition of unity \eqref{eq:3.1.10}, we represent the operator $\Delta \cL$ as
$$
\Delta \cL=\Delta \cL\,\zeta_{0}+\sum\limits_{r=1}^{\exnum}\Delta \cL\,\zeta_{r}.
$$
The operator $\Delta \cL\,\zeta_{0}$ is finite-dimensional, since the support of the function $\Delta \cL\,(\zeta_{0}u)$ is contained in a finite set independent of $u$. By the estimate \eqref{eq:3.1.16}, for large $t$ each of the operators $\Delta \cL\, \zeta_r: H_{\nu}(V) \rightarrow \cH_{\nu}(\mathring{V}, \partial V)$ has small norm. Thus the operator $\Delta \cL$ is approximated by finite-dimensional operators and hence is compact.

The compactness of the operator $\Delta \cL$ allows us to construct a right regularizer for $\mathcal{L}_{\nu}^{(0)}(\mu)$ rather than for $\mathcal{L}_{\nu}(\mu)$. Indeed, if 
\begin{equation}\label{eq:3.1.22}
	\mathcal{L}_{\nu}^{(0)}(\mu)\,\cR_{\nu}^{(0)}(\mu)=I+\mathcal{K}^{(0)},
\end{equation}
where $\mathcal{K}^{(0)}$ is a compact operator, then
\begin{equation}\label{eq:3.1.23}
	\mathcal{L}_{\nu}(\mu)\,\cR_{\nu}^{(0)}(\mu)=I+\mathcal{K}
\end{equation}
with the compact operator $\mathcal{K}:=\mathcal{K}^{(0)}+\Delta \cL\, \cR_{\nu}^{(0)}(\mu)$.

We proceed to the construction of the regularizer $\cR_{\nu}^{(0)}(\mu)$. Let us consider the problem
\begin{align}\label{eq:3.1.24}
	(A^{(0)}-\mu)u(x)&=f(x),\quad x\in \mathring{V},\\
	u(x)&=g(x), \quad x \in \partial V,
\end{align}
with right-hand side $\{f,g\}\in \cH_{\nu}(\mathring{V}, \partial V)$. Let, as before, $\zeta_{0},\ldots,\zeta_{\exnum}$ be the partition of unity \eqref{eq:3.1.10}, where the number $t$ is chosen so that
\begin{equation}\label{eq:3.1.25}
	\begin{gathered}
		t>\max\limits_{r=1,\ldots,\exnum}t_{r}
	\end{gathered}
\end{equation}
and $t_r$ is the number introduced after \eqref{eq:3.1.15}.
We represent the function $f$ as
$$
f=\zeta_{0}f+\sum\limits_{r=1}^{\exnum}\zeta_{r}f,
$$
and do the same for the function $g$.
The supports of the functions $\zeta_{r}f$, $\zeta_{r}g$ are contained in the semi-cylinder $V\cap V^{+}_{r}$; let us transfer these functions from $V$ to $V_{r}$ and keep the old notation. On $V_{r}$ consider the model problem \eqref{eq:3.1.5_Inn}, \eqref{eq:3.1.5_Bound} with right-hand side $\{\zeta_{r}f, \zeta_{r}g\}$. Since the interval $\mathbb{T}+i\nu$ is free of poles of the resolvent $\lambda\mapsto R^D_r(\lambda,\mu)$, by Theorem \ref{th:2.3.7} there exists a unique solution $u_{r}$ of this problem. Keeping the old notation, we transfer the function $\zeta_r$ in \eqref{eq:3.1.10} from $V$ into $V_r$:
\begin{equation}\label{eq:3.1.26}
	\zeta_r(x):=\left\{\begin{array}{ll}
		1,&\;x=(x_{r}^{*},n_{r}),\;n_{r}\geqslant t,\\
		0,&\;x=(x_{r}^{*},n_{r}),\;n_{r}< t.
	\end{array}\right.
\end{equation}
For each $r=1, \dots, \exnum$ let us transfer the function $\zeta_r u_r$ from the cylinder $V_r$ into the waveguide $V$, keeping the previous notation. Set
\begin{equation}\label{eq:3.1.27}
	\cR_{\nu}^{(0)}(\mu)\{f,g\}=\sum\limits_{r=1}^{\exnum}\zeta_r u_{r}.
\end{equation}
We now check that this operator is a right regularizer for $\mathcal{L}_{\nu}^{(0)}(\mu)$. For arbitrary $m\in \bN$ set
$$
V_r^+(m):=\{(x_r^*, n_r) \in V_r^+: x_r^*\in V_r^*, n_r\geq m\}.
$$
Vertices in $V_r^+(t)$ can be connected by edges only to vertices in 
$V_r^+(t-t_r)$; therefore the function $(A^{(0)}-\mu)(\zeta_r u_r)$ vanishes on $V\setminus V_r^+(t-t_r)$. By Proposition \ref{pr:2.3.14} we have $(A^{(0)}-\mu)(\zeta_r u_r)=(A^{(0)}-\mu) u_r$ on $V\cap V_r^+(t+t_r)$.

The coefficients of the operators $A^{(0)}$ and $A_r$ coincide in the semi-cylinder $\Pi_r^+$ (see the beginning of Section \ref{subsec:3.1}); therefore $A^{(0)}u=A_r u$ in $V\cap V_r^+(t_{r})$ for any function $u$. From the above and the identity \eqref{eq:3.1.5_Inn} it follows that 
\begin{align}
	(A^{(0)}-\mu)(\zeta_r u_r)&=(A_r-\mu) u_r=\zeta_r f=f \;\text{ on } \mathring{V}\cap V_r^+(t+t_r),\\
	(A^{(0)}-\mu)(\zeta_r u_r)&=0 \;\text{ on } \mathring{V}\setminus V_r^+(t-t_r).
\end{align}
Summing these identities over $r=1, \dots, \exnum$, we obtain
\begin{align}
	(A^{(0)}-\mu)\cR_{\nu}^{(0)}(\mu)\{f,g\}=f \;\text{ on } \bigcup\limits_{r=1}^\exnum \left(\mathring{V}\cap V_r^+(t+t_r)\right).
\end{align}
Reasoning similarly, we obtain
\begin{align}
	\cR_{\nu}^{(0)}(\mu)\{f,g\}=g \;\text{ on } \bigcup\limits_{r=1}^\exnum \left(\partial V\cap V_r^+(t)\right).
\end{align}
Then 
$$
\cL_\nu^{(0)} \,\cR_{\nu}^{(0)}(\mu)\{f,g\}=\{f,g\}+\{\Delta f, \Delta g\}.
$$
Here the defect $\{\Delta f, \Delta g\}$ has support lying in the finite set
$$
V\setminus\bigcup\limits_{r=1}^\exnum \left(V\cap V_r^+(t+t_r)\right),
$$
independent of the pair $\{f,g\}$. Therefore the operator $\{f,g\} \mapsto \{\Delta f, \Delta g\}$ is finite-dimensional and hence compact.

\paragraph{4. Finite-dimensionality of the cokernel.}

It suffices to note that $\dim\coker(I+\mathcal{K})<\infty$ for every compact operator $\mathcal{K}$, and to use the identity $\mathcal{L}_{\nu}(\mu)\cR_{\nu}^{(0)}(\mu)=I+\mathcal{K}$, proved in the previous step.

\paragraph{5. Necessity of the conditions of the theorem.}

Assume, as before, that $\mu$ is distinct from all flat bands of the problems \eqref{eq:3.1.5_Inn}, \eqref{eq:3.1.5_Bound} for $r=1,\dots,d$. Suppose that the operator \eqref{eq:3.1.7} is Fredholm. We show that then the interval $\mathbb{T}+i\nu$ is free of poles of the resolvent $\lambda\mapsto R^D_{r}(\lambda,\mu)$, $r=1,\ldots,\exnum$. Let, as in the second step of the proof, $\mathcal{H}$ be a direct complement in $H_{\nu}(V)$ to the finite-dimensional subspace $\ker\mathcal{L}_{\nu}(\mu)$. By the closedness of the range $\rran\mathcal{L}_{\nu}(\mu)$, the mapping $\mathcal{L}_{\nu}(\mu):\mathcal{H}\rightarrow\mathrm{ran}\mathcal{L}_{\nu}(\mu)$ is an isomorphism. Let $\cR_{\nu}(\mu)$ stand for the inverse operator for this mapping. From the inequality $\dim\ker\mathcal{L}_{\nu}(\mu)<\infty$ it follows that on $H_{\nu}(V)$ the identity $\cR_{\nu}(\mu)\mathcal{L}_{\nu}(\mu)=I+\mathcal{K}$ holds, where $\mathcal{K}$ is a finite-dimensional operator. This yields the estimate
\begin{equation}\label{eq:3.1.32}
	\|u;H_{\nu}(V)\|\leqslant c\left(\|(A-\mu)u;H_{\nu}(\mathring{V})\|+\|u;H_{\nu}(\partial V)\|+\|\mathcal{K}u;H_{\nu}(V)\|\right).
\end{equation}
Suppose now that the interval $\mathbb{T}+i\nu$ contains a pole $\lambda_{0}$ of the resolvent $R^D_{r}(\lambda,\mu)$ for some $r$; hereafter the index $r$ is omitted for brevity and no confusion arises. Then for the operator $A^D(\lambda)$ there exist an eigenvalue $\mu(\lambda)$ and an eigenvector $\psi(\lambda)$ such that $\mu(\lambda_{0})=\mu$ and $(A^D(\lambda_{0})-\mu)\psi(\lambda_{0})=0$. Let $v$ be a solution of the form \eqref{eq:2.3.13_Dir} to the homogeneous problem \eqref{eq:2.2.1_Dir}:
$$
v(x^{*},n):=\exp(i\lambda_{0} n)P^*\psi(x^{*},\lambda_{0}).
$$
For $N \in \bN$ let $\chi_N$ be the cutoff function on $V_r$: 
\begin{align}\label{eq:chi_cutoff}
	\chi_N(x^*,n)=\left\{\begin{array}{ll}
		1,& t_r+N^2\leq n< t_r+(N+1)^2,\\
		0,&\text{otherwise}.
	\end{array}\right.
\end{align}
We introduce on $V\cap V^{+}_{r}$ the function $\chi_N v$ and extend it by zero to the remaining part of $V$, keeping the previous notation. The value $\exp(\nu n)|v(x^{*},n)|$ is independent of $n$; hence $\|\chi_{N} v;H_{\nu}(V)\|=c(2N+1)^{1/2}$ with some constant $c>0$. On $\partial V$ the function $\chi_N v$ is zero. Let us prove the formula
\begin{align}\label{eq:Achiv_est}
	\|(A-\mu)(\chi_{N} v);H_{\nu}(\mathring{V})\|=o(N^{1/2})
\end{align}
as $N\to \infty$. We have
$$
(A-\mu)(\chi_{N} v)=(A^{(0)}-\mu)(\chi_{N} v)+(\Delta A)(\chi_{N} v).
$$
We estimate each term in the right-hand side in isolation. 
The stabilization conditions \eqref{eq:3.1.4} imply
$$
\|\Delta A(\chi_{N} v);H_{\nu}(\mathring{V})\|=o(1)\cdot\|\chi_{N} v;H_{\nu}(V)\|=o(N^{1/2}).
$$
The coefficients of the operators $A^{(0)}$ and $A_r$ coincide in the semi-cylinder $\Pi_r^+$, and the function $v$ satisfies the identity $(A_r\, v)(x)=0$ for $x\in \mathring{V}_r$. Then by Proposition \ref{pr:2.3.14} the support of the function $(A^{(0)}-\mu)(\chi_N v)$ lies in the union of the sets $$\{(x^*,n)\in \mathring{V}\cap V_r^+: N^2\leq n< N^2+2t_r\}, \quad \{(x^*,n)\in \mathring{V}\cap V_r^+: (N+1)^2\leq n< (N+1)^2+2t_r\}.
$$
The sets have finite cardinalities. Therefore,  $\|(A^{(0)}-\mu)(\chi_{N} v);H_{\nu}(\mathring{V})\|=O(1)$ as $N\to \infty$. The estimate \eqref{eq:Achiv_est} is proved.

Set 
$$
w_N:=\chi_{N} v/\|\chi_{N} v;H_{\nu}(V)\|.
$$ 
The formulas obtained above imply that $\|(A-\mu)w_N;H_{\nu}(\mathring{V})\|\to 0$ as $N \to \infty$. By \eqref{eq:chi_cutoff} the supports of the functions $w_N$ for different $N$ are pairwise disjoint, and the sequence $\{w_N\}$ is orthonormal in $H_{\nu}(V)$. Therefore $w_N$ converges weakly to zero as $N\to \infty$. By the compactness of the operator $\cK$, the expression $\cK w_{N}$ converges strongly to zero. Consequently, the sequence $w_N$ contradicts the estimate \eqref{eq:3.1.32}.

The proof of Theorem \ref{th:3.1.1} is complete.

\begin{rem}\label{rm:3.1.3}
	The space $\ker\mathcal{L}_{\nu}(\mu)$ remains finite-dimensional even in the case where the interval $\mathbb{T}+i\nu$ contains poles of the resolvents $R^D_r(\lambda,\mu)$. Indeed, let us replace the number $\nu$ by a number $\gamma<\nu$ such that the interval $\mathbb{T}+i\gamma$ has no poles. Since $H_{\nu}(V)\subset H_{\gamma}(V)$, we have $\ker\mathcal{L}_{\nu}(\mu)\subset \ker\mathcal{L}_{\gamma}(\mu)$. Moreover, $\dim\ker\mathcal{L}_{\gamma}(\mu)<\infty$ by Theorem {\rm\ref{th:3.1.1}}. Similarly one verifies that $\dim\coker \mathcal{L}_{\nu}(\mu)<\infty$ for $\nu\in\mathbb{R}$. Under the assumptions of Theorem \ref{th:3.1.1}, the presence of poles of the resolvents $R^D_{r}(\lambda,\mu)$ on the interval $\mathbb{T}+i\nu$ is equivalent to the non-closedness of  $\rran\mathcal{L}_{\nu}(\mu)$.
\end{rem}

\subsection{Asymptotics of solutions. The index of the operator \eqref{eq:3.1.7}}

To derive asymptotic formulas, we have to make more restrictive assumptions on the stabilization of the coefficients than those in \eqref{eq:3.1.4}. We assume that the coefficients stabilize exponentially, i.e., for all $r=1,\ldots\exnum$, $e_{r}^{*}\in E_{r}^{*}$, and $x_{r}^{*}\in V_{r}^{*}$, as $n_{r}\rightarrow+\infty$, the formulas
\begin{equation}\label{eq:3.2.1}
	\begin{gathered}
		|a(e_{r}^{*},n_{r})-a_r(e_{r}^{*})|=O(e^{-\delta n_{r}}),\\
		|q(x_{r}^{*},n_{r})-q_r(x_{r}^{*})|=O(e^{-\delta n_{r}})
	\end{gathered}
\end{equation}
hold with some $\delta>0$.
Let the spectral parameter $\mu$ coincide with neither flat bands nor thresholds of the operators $A_r(\lambda)$ for any $r$. Let also $\beta$ be a real number such that $0<\beta<\delta$, and let the rectangle $\{\lambda\in\mathbb{C}:|\mathrm{Im}\lambda|\leqslant\beta, \, 0\leqslant\mathrm{Re}\lambda<2\pi\}$ contain only real poles of the resolvents $\lambda\mapsto R_r^D(\lambda,\mu)$. Let us number these poles by the index $k=1,\ldots,\varkappa_r$ with account of multiplicity (see the explanation before formula {\rm\eqref{eq:2.3.13}}). Set
\begin{equation}\label{eq:3.3.1}
	\varkappa:=\sum\limits_{r=1}^{\exnum}\varkappa_{r}.
\end{equation}
An asymptotics of the solutions of the problem \eqref{eq:3.1.1} is established in isolation for each cylindrical outlet. In the following proposition we assume that such an outlet is fixed and has index $r$. Let $\zeta_{r}$ be the cutoff \eqref{eq:3.1.10} with $t$ fixed and satisfying the inequality \eqref{eq:3.1.25}.

\begin{thm}\label{th:3.2.1}
	Let $u$ be a solution of the problem \eqref{eq:3.1.1} satisfying the condition $\zeta_{r} u\in H_{-\beta}(V)$, and let the right-hand side $\{f,g\}$ satisfy the inclusion $\{\zeta_{r} f, \zeta_{r} g\}\in \cH_{\beta}(\mathring{V}, \partial V)$. Then the formula
	\begin{equation}\label{eq:3.2.2}
		(\zeta_r u) (x_{r}^{*},n_{r})= v_r(x_{r}^{*},n_{r})+\sum\limits_{k=1}^{\varkappa_{r}}c_{r,k}v_{r,k}(x_{r}^{*},n_{r})
	\end{equation}
	holds on $V\cap V^{+}_{r}$, where $v_{r,k}$ are the functions {\rm\eqref{eq:2.3.13_Dir}} for the cylinder $\Pi_{r}$, $c_{r,k}$ are constants, and $\zeta_{r} v_r\in H_{\beta}(V)$.
\end{thm}

\begin{proof}
	The proof relies on Theorem \ref{th:2.3.8_Dir}, where an asymptotics of the solution to problem in a cylinder is described. Analysis similar to that in Step 1 of Theorem \ref{th:3.1.1} shows that the function $\zeta_r u$ satisfies the system \eqref{eq:3.1.12_Inn}, \eqref{eq:3.1.12_Bound}:
	\begin{align}
		(A^{(0)}-\mu)(\zeta_{r}u)&=F \text{ on } \mathring{V},\label{eq:3.1.12_Inn2}\\
		\zeta_{r}u&=\zeta_{r}g \text{ on } \partial V,\label{eq:3.1.12_Bound2}
	\end{align}
	where $F:=\zeta_{r}f+[A,\zeta_{r}]u+\Delta A(\zeta_{r}u)$ is a function with support in $V\cap V_r^+$. Let us Transfer all the functions in \eqref{eq:3.1.12_Inn2}, \eqref{eq:3.1.12_Bound2} from $V$ to $V_r$; then this system can be considered as a problem in the cylinder $\Pi_{r}$. 
	
	We prove the asymptotic formula \eqref{eq:3.2.2} in two steps. In the first step we establish the inclusion $\zeta_r u\in H_{\beta-\delta}(V_{r})$. Since the support of the function $\zeta_r u$ is contained in $V_r^+$, the inclusion $\zeta_r u\in H_{-\beta}(V_{r})$ implies $\zeta_r u\in H_{\nu}(V_{r})$ for $\nu\leq-\beta$; the same considerations are applied below to other functions. Therefore, if $\delta\geq 2\beta$, the inclusion $\zeta_r u\in H_{\beta-\delta}(V_{r})$ is automatic and the first step of the proof can be skipped.
	
	Let us assume that $0<\beta<\delta<2\beta$. We first prove the inclusion $\{F, \zeta_{r} g\}\in \cH_{-\beta}(\mathring{V}_r, \partial V_r)\cap \cH_{\beta-\delta}(\mathring{V}_r, \partial V_r)$. By Proposition \ref{pr:2.3.14} the expression $[A,\zeta_r]u$ has finite support and hence belongs to $H_{\nu}(\mathring{V}_{r})$ for any $\nu\in \bR$. By assumption, $\{\zeta_{r} f, \zeta_{r} g\}\in \cH_{\beta}(\mathring{V}, \partial V)$ and hence $\{\zeta_{r} f, \zeta_{r} g\}\in \cH_{\nu}(\mathring{V}_r, \partial V_r)$ for any $\nu\leq\beta$. Finally, by the inclusion $\zeta_r u\in H_{-\beta}(V_{r})$ and the stabilization conditions \eqref{eq:3.2.1}, the function $\Delta A(\zeta_r u)$ belongs to $H_{\nu}(\mathring{V}_{r})$ for $\nu\leq\delta-\beta$. Thus we have proved the inclusion $\{F, \zeta_{r} g\}\in \cH_{\nu}(\mathring{V}_r, \partial V_r)$ for $\nu\leq \delta-\beta$, in particular, for $\nu=\beta-\delta$. Since the rectangle $\{\lambda\in\mathbb{C}:-\beta\leq\mathrm{Im}\lambda\leq \beta-\delta, \, 0\leqslant\mathrm{Re}\lambda<2\pi\}$ is free of poles of the resolvent $\lambda\mapsto R_r^D(\lambda,\mu)$, by Theorem \ref{th:2.3.8_Dir} (see Remark \ref{rm:2.3.11}) we have $\zeta_r u\in H_{\beta-\delta}(V_{r})$.
	
	In the second step, using the stabilization conditions \eqref{eq:3.2.1} and the inclusion $\zeta_r u\in H_{\beta-\delta}(V_{r})$ proved above, we derive $\Delta A(\zeta_r u)\in H_{\beta}(\mathring{V}_{r})$. Then from the inclusions in the previous paragraph yield $\{F, \zeta_{r} g\}\in \cH_{-\beta}(\mathring{V}_r, \partial V_r)\cap \cH_{\beta}(\mathring{V}_r, \partial V_r)$. Now, applying Theorem \ref{th:2.3.8_Dir} for the exponents $\beta$ and $-\beta$, we obtain the formula \eqref{eq:3.2.2}.
\end{proof}

\begin{rem}\label{rm:3.2.2}
	The formula \eqref{eq:2.3.16_Dir} from Theorem {\rm\ref{th:2.3.8_Dir}} does not allow one to express the coefficients $c_{k}$ as functionals of $\{f,g\}$, because the right-hand side of the equation \eqref{eq:3.1.12_Inn2} in its turn depends on $u$. The formulas for $c_{k}(f,g)$ will be found below in Theorem {\rm\ref{th:3.3.7}} under additional assumptions.
\end{rem}

In the justification of the main theorem \ref{th:3.3.7} we shall need analogues of the statements established in the book \cite{NazPlam} (Chapter 4, §3 and Chapter 5, §3) for waveguides described by boundary value problems for partial differential equations. However, the proofs of these statements do not use the specifics of the problem considered; they rely on the analogues of Theorems \ref{th:3.1.1} and \ref{th:3.2.1} only. This allows one to transfer them to discrete waveguides without essential changes. Therefore the statements \ref{th:3.3.3}-\ref{th:3.3.6} are given without proofs or with brief explanations.

Recall that the index of a Fredholm operator $\mathcal{L}$ is the number
$$
\mathrm{Ind}\mathcal{L}=\dim\ker \mathcal{L}-\dim\coker \mathcal{L}. 
$$
The next theorem is an analogue of Theorem 4.3.3 in \cite{NazPlam}.
\begin{thm}\label{th:3.3.3}
	Let $\cL_\beta(\mu)$ be the operator \eqref{eq:3.1.7}. Then there holds the formula
	$$
	\mathrm{Ind}\mathcal{L}_{-\beta}(\mu)=\mathrm{Ind}\mathcal{L}_{\beta}(\mu)+\varkappa
	$$
	holds.
\end{thm}

We obtain an important corollary of this formula. Let us assume that the conditions of Theorem \ref{th:3.1.1} are satisfied and, consequently, $\mathcal{L}_{\nu}(\mu):H_\nu(V)\to \cH_{\nu}(\mathring{V}, \partial V)$ is a Fredholm operator. Then system \eqref{eq:3.1.1} with right-hand side  $\{f,g\} \in \cH_{\nu}(\mathring{V}, \partial V)$ has a solution $u \in H_\nu(V)$ if and only if $\{f,g\}$ satisfies a finite number of conditions, which is equal to the dimension of the cokernel. A description of such conditions is given in the next two propositions. The Banach space $\cH_{\nu}(\mathring{V}, \partial V)$ is reflexive, and $\cH_{\nu}(\mathring{V}, \partial V)^*$ is isomorphic to $\cH_{-\nu}(\mathring{V}, \partial V)$ with respect to the pairing
$$(\{f,g\}, \{\Psi, \Phi\})_{\cH(\mathring{V}, \partial V)}:= (f,\Psi)_{H(\mathring{V})}+(g, \Phi)_{H(\partial V)},$$
obtained by continuous extension of the corresponding inner products to pairs $\{f,g\} \in \cH_{\nu}(\mathring{V}, \partial V)$ and $\{\Psi, \Phi\} \in \cH_{-\nu}(\mathring{V}, \partial V)$.

\begin{prop}\label{pr:coker}
	Let the pair $\{\Psi, \Phi\} \in \cH_{-\nu}(\mathring{V}, \partial V)$ satisfy the identity
	\begin{align}\label{eq:coker}
		(\mathcal{L}_{\nu}(\mu)u, \{\Psi, \Phi\})_{\cH(\mathring{V}, \partial V)}=0
	\end{align}
	for all  $u \in H_\nu(V)$. 
	Then $P^*\Psi$ belongs to the space $\ker \cL_{-\nu}(\mu)$, and the identity
	\begin{align}\label{eq:coker3}
		\Phi=-B^N P^*\Psi
	\end{align}
	holds. Here $B^N$ is the boundary operator \eqref{eq:1.3.3} and $P^*$ stands for the operator of continuation by zero in \eqref{eq:oper}.
\end{prop}
\begin{proof}
	In expanded form the identity \eqref{eq:coker} reads
	\begin{align}\label{eq:coker1}
		((A-\mu) u,\Psi)_{H(\mathring{V})}+(u, \Phi)_{H(\partial V)}=0.
	\end{align}
	Green's formula \eqref{eq:3.1.3} extends by continuity to the set of pairs $(u,v) \in H_\nu(V)\times H_{-\nu}(V)$. 
	We first choose a function $u \in H_\nu(V)$ that vanishes on $\partial V$. Then the second term in \eqref{eq:coker1} vanishes, and by Green's formula \eqref{eq:3.1.3} we obtain
	$$
	0=((A-\mu) u,\Psi)_{H(\mathring{V})}=(u,(A-\mu)P^*\Psi)_{H(\mathring{V})}. 
	$$
	Since $u \in H_\nu(V)$ is arbitrary on $\mathring{V}$, we have 
	\begin{align}
		(A-\mu)P^*\Psi=0 \quad \text{on}\quad\mathring{V}.
	\end{align}
	Now suppose that the function $u \in H_\nu(V)$ is arbitrary. Then by Green's formula \eqref{eq:3.1.3}, 
	$$
	0=(u,(A-\mu)P^*\Psi)_{H(\mathring{V})} +(-B^N u,  P^*\Psi)_{H(\partial V)}=((A-\mu)u,P^*\Psi)_{H(\mathring{V})}+(u, -B^N P^*\Psi)_{H(\partial V)}.
	$$
	Comparing the right-hand side of this identity with \eqref{eq:coker1}, we obtain
	$$
	(u, -B^N P^*\Psi-\Phi)_{H(\partial V)}=0.
	$$
	Since the function $u$ is arbitrary on $\partial V$, we arrive to \eqref{eq:coker3}.
\end{proof}
From Proposition \ref{pr:coker}, the Fredholmness of the operator $\cL_{\nu}(\mu)$, and the reflexivity of the space $\cH_{\nu}(\mathring{V}, \partial V)$, the following proposition holds.
\begin{prop}\label{pr:coker2}
	The system \eqref{eq:3.1.1} with right-hand side $\{f,g\}\in \cH_{\nu}(\mathring{V}, \partial V)$ is solvable in $H_\nu(V)$ if and only if, for all $z \in \ker \cL_{-\nu}$, the identity
	\begin{align}\label{eq:coker2}
		(f,z)_{H(\mathring{V})}+(g, -B^N z)_{H(\partial V)}=0
	\end{align}
	holds.
\end{prop}
By Proposition \ref{pr:coker2} we have $\dim \coker\mathcal{L}_{\pm\beta}(\mu)=\dim \ker\mathcal{L}_{\mp\beta}(\mu)$. Therefore
\begin{align}\label{eq:vark_Ups}
	\varkappa=\mathrm{Ind}\mathcal{L}_{-\beta}(\mu)-\mathrm{Ind}\mathcal{L}_{\beta}(\mu)=2(\dim\ker\mathcal{L}_{-\beta}(\mu)-\dim\ker\mathcal{L}_{\beta}(\mu))=:2\Upsilon.
\end{align}

\subsection{Waves, the continuous spectrum eigenfunctions, and the scattering matrix}
For each cylinder $\Pi_r$ related to the cylindrical outlet $G\cap \Pi_r^+$, $r=1, \dots, \exnum$, we construct in $V_r$ the waves $\{w_{k,r}\}_{k=1}^{\varkappa_r}$ of the form \eqref{eq:2.3.25}, transfer such functions from $V_r$ into $V$ and call them \emph{waves in the waveguide} $G$. The resulting set of waves is numbered by a single index independently of the outlet number and denoted by $\{w_{k}\}_{k=1}^{\varkappa}$. The linear span $\bW$ of these functions is called the wave space in the waveguide. 
We are to choose a basis in $\bW$ that consists of incoming and outgoing waves. To this end, for $t$ satisfying \eqref{eq:3.1.25}, set
$$
V_{r}^{+}(t)=\{(x_{r}^{*},n_{r})\in V_{r}^{+}: n_{r}\geqslant t\},\quad V(t)=V\setminus\bigcup\limits_{r=1}^{\exnum
}V_{r}^{+}(t).
$$
For waves $u,v$ in the wave space $\bW$, introduce the sesquilinear form
$$
q_{t}(u,v):=((A-\mu)u,v)_{H(\mathring{V}(t))}+(u,-B^N  v)_{H(\partial V(t))}-(u,(A-\mu)v)_{H(\mathring{V}(t))}-(-B^N u, v)_{H(\partial V(t))},
$$
where $\mathring{V}(t):=\mathring{V}\cap V(t)$ and $\partial V(t):=\partial V\cap V(t)$.
Its values depend on $t$, but the following proposition holds.
\begin{prop}\label{pr:3.3.4}
	For any $u,v\in\bW$ there exists the finite limit
	\begin{equation}\label{eq:3.3.2}
		q(u,v):=\lim\limits_{t\rightarrow\infty}q_{t}(u,v).
	\end{equation}
	Moreover, if the supports of the functions $u$ and $v$ lie in $V\cap V^{+}_{r}$ and $V\cap V^{+}_{s}$, respectively, then
	\begin{equation}\label{eq:3.3.3}
		q(u,v)=0\;\text{if}\;r\neq s,\qquad q(u,v)=q_{\Pi_r}(\tilde{u},\tilde{v})\;\text{if}\;r=s.
	\end{equation}
	Here $q_{\Pi_r}(\cdot,\cdot)$ is the sesquilinear form \eqref{eq:2.3.21} for the cylinder $\Pi_{r}$, and $\tilde{u},\;\tilde{v}$ are the waves in the cylinder $\Pi_{r}$ obtained by transferring $u$ and $v$ from $V\cap V^{+}_{r}$ into $V_r$.
\end{prop}

A wave $u\in\bW$ is called outgoing (incoming) if the number $iq(u,v)$ is positive (negative). Waves $u$ and $v$ are called orthogonal if $iq(u,v)=0$. 
The form $q(\cdot,\cdot)$ can be extended to $(\bW\dotplus H_{\beta}(V))\times(\bW\dotplus H_{\beta}(V))$. Indeed, formula \eqref{eq:3.1.3} extends by continuity to $H_{-\beta}(V)\times H_{\beta}(V)$. Therefore, if at least one of the functions $u,v\in \bW\dotplus H_{\beta}(V)\subset H_{-\beta}(V)$ belongs to $H_{\beta}(V)$, then $q(u,v)=0$. Consequently, for any waves $u,\;v\in\bW$ and any functions $u_{0},\;v_{0}\in H_{\beta}(V)$, the identity
\begin{equation}\label{eq:3.3.4}
	iq(u+u_{0},v+v_{0})=iq(u,v)
\end{equation}
holds.

\begin{prop}\label{pr:3.3.5}
{\rm 1.} In the space $\bW$ there exists a basis $u_{1}^{+},\ldots,u_{\Upsilon}^{+},\;u_{1}^{-},\ldots,u_{\Upsilon}^{-}$ satisfying the orthogonality and normalization relations
		\begin{equation}\label{eq:3.3.5}
			iq(u_{j}^{\pm},u_{k}^{\mp})=0,\quad iq(u_{j}^{\pm},u_{k}^{\pm})=\mp\delta_{jk},\quad j,k=1,\ldots,\Upsilon.
		\end{equation}
		Thus the waves $u_{j}^{+}$ are incoming, and the waves $u_{j}^{-}$ are outgoing.\\
{\rm 2.} Every orthogonal basis in the space $\bW$ contains $\Upsilon$ incoming and $\Upsilon$ outgoing waves.
\end{prop}

\begin{proof}
	In each cylinder the waves \eqref{eq:2.3.25} were constructed. With respect to the form \eqref{eq:2.3.21}, they are incoming or outgoing orthogonal waves. By \eqref{eq:3.3.3}, the waves in the waveguide obtained by transferring the waves \eqref{eq:2.3.25} from all cylinders are orthogonal and are either incoming or outgoing. By \eqref{eq:vark_Ups}, the number of such waves equals $2\Upsilon$, where $\Upsilon=\dim\ker\mathcal{L}_{-\beta}(\mu)-\dim\ker\mathcal{L}_{\beta}(\mu)$. It remains to check that in such a set the numbers of incoming and outgoing waves are equal. This fact and item 2 are verified in the same way as in Theorem 5.3.2 in \cite{NazPlam}.
\end{proof}

The following theorem is an analogue of Theorem 5.3.4 in \cite{NazPlam}.
\begin{thm}\label{th:3.3.6}
	In the quotient space ${\ker\mathcal{L}_{-\beta}(\mu)}/{\ker\mathcal{L}_{\beta}(\mu)}$ there exists a unique basis with representatives $Y_{1}^{+},\ldots,Y_{\Upsilon}^{+}$ satisfying the relations
	\begin{equation}\label{eq:3.3.6}
		\begin{gathered}
			Y_{j}^{+}-u_{j}^{+}-\sum\limits_{l=1}^{\Upsilon}u_{l}^{-}S_{lj}\in H_{\beta}(V),\quad j=1,\ldots,\Upsilon.
		\end{gathered}
	\end{equation}
	The matrix $S$ with elements $S_{lj}$ is unitary: $S^{*}=S^{-1}$.
\end{thm}
Let us clarify the statement of Theorem \ref{th:3.3.6}:

1. The matrix $S$ in Theorem \ref{th:3.3.6} is called the scattering matrix. The functions $Y_j^+$ can be regarded as the result of scattering of the incoming wave $u_j^+$ into the outgoing waves $u_l^-$.

2. Recall that an eigenfunction of the problem \eqref{eq:3.1.1} is a solution to the corresponding homogeneous problem from the class $H(V)$. If such a solution exists, the value $\mu$ is called an eigenvalue of the problem \eqref{eq:3.1.1}. 

It follows from Theorem \ref{th:3.3.6} that the space $\ker\cL_\beta(\mu)$ is the space of eigenfunctions of the problem \eqref{eq:3.1.1} corresponding to the number $\mu$. Indeed, every eigenfunction $z \in H(V)$ belongs to $H_{-\beta}(V)$ and is therefore a linear combination of the functions $Y_j^+$ up to a remainder in $H_\beta(V)$.
The waves $u_j^\pm$ are linearly independent modulo $H(V)$, i.e. their linear combination 
$$\sum_{j=1}^{\Upsilon}\left(c_j^+ u_j^+ + c_j^- u_j^-\right)$$
belongs to $H(V)$ only when $c_j^\pm=0$, $j=1, \dots, \Upsilon$. Therefore, by \eqref{eq:3.3.6}, the function $z$ belongs to $\ker\cL_{\beta}(\mu)$. 

3. An eigenfunction of the continuous spectrum is a solution $Y$ to the homogeneous problem \eqref{eq:3.1.1} that does not belong to $H(V)$, but admits the estimate $|Y(x)|\leq {\rm const} (1+|x|)^N$ with some $N<\infty$. Here $|x|$ is the function from formula \eqref{eq:3.1.6}. The value $\mu$ is called a point of the continuous spectrum of the problem \eqref{eq:3.1.1} if  the continuous spectrum eigenfunctions exist for this $\mu$.

Theorem \ref{th:3.3.6} implies that the set $\ker\cL_{-\beta}(\mu)\setminus\ker\cL_\beta(\mu)$ is the set of the continuous spectrum eigenfunctions for the point $\mu$ (this is true even in the case when $\mu$ is a threshold, but we don't discuss it in the paper). Indeed, since the number $\mu$ is different from the thresholds of the problem \eqref{eq:3.1.1}, the waves $u_j^\pm$ are bounded functions of the point $x\in V$ (see \eqref{eq:2.3.25_Dir}).  Therefore, by Theorem \ref{th:3.3.6}, the functions from $\ker \cL_{-\beta}(\mu)$ are bounded.

4. Let $A^D:H(\mathring{V})\to H(\mathring{V})$ be the self-adjoint operator \eqref{eq:oper}. According to the standard theory of self-adjoint operators in Hilbert spaces, the point $\mu$ is called a point of the continuous spectrum of the operator $A^D$ if the range of the operator $A^D-\mu I$ is not closed in $H(\mathring{V})$. From Theorems \ref{th:3.1.1} and \ref{th:3.3.6} 
it follows that the number $\mu$ belongs to the continuous spectrum of the operator $A^D$ if and only if $\mu$ is a point of the continuous spectrum of the problem \eqref{eq:3.1.1}, i.e., if eigenfunctions of the continuous spectrum exist. 
\begin{prop}
	The continuous spectrum of the operator $A^D$ coincides with the union of the ranges of all band functions $\bR \ni \lambda \mapsto\mu_j(\lambda)$ from \eqref{eq_A_lam_eigenfun} for the problems in the cylinders $\Pi_1, \dots, \Pi_\exnum$. Therefore, the spectrum has a band structure.
\end{prop}

\subsection{The radiation principle}

``Radiation conditions'' consist in specifying the type of the asymptotics of the sought solution. We will seek for the solutions that have asymptotics with leading term consisting of outgoing waves. Such a radiation condition is called intrinsic. For the next theorem we need continuous spectrum eigenfunctions that differ from those in Theorem \ref{th:3.3.6}. Set
\begin{equation}\label{eq:3.3.8}
	Y_{j}^{-}:=\sum\limits_{l=1}^{\Upsilon}Y_{l}^{+}(S^{-1})_{lj}, \quad j=1,\ldots,\Upsilon.
\end{equation}
Formulas \eqref{eq:3.3.6} and \eqref{eq:3.3.8} imply that
\begin{equation}\label{eq:3.3.13}
	\begin{gathered}
		Y_{j}^{-}-u_{j}^{-}-\sum\limits_{l=1}^{\Upsilon}u_{l}^{+}(S^{-1})_{lj}\in H_{\beta}(V),\quad j=1,\ldots,\Upsilon.
	\end{gathered}
\end{equation}
The functions $Y_{j}^{-}$ are similar to $Y_{j}^{+}$, but with the incoming waves replaced by the outgoing ones, and vice versa.
\begin{thm}\label{th:3.3.7}
	Let $\{z_{1},\ldots,z_{\Theta}\}$ be a basis in the space $\ker\mathcal{L}_{\beta}(\mu)$. Assume that $\{f,g\}$ belong to $\cH_{\beta}(\mathring{V}, \partial V)$ and that the orthogonality conditions 
	\begin{align}\label{eq:sol_cond}
		(f,z_{j})_{H(\mathring{V})}+(g,-B^N z_{j})_{H(\partial V)}=0, \quad j=1,\ldots,\Theta,
	\end{align}
	hold. Then\\
	{\rm 1.} There exists a solution $u$ of the problem \eqref{eq:3.1.1} satisfying the radiation conditions
	\begin{equation}\label{eq:3.3.7}
		v:=u-c_{1}u_{1}^{-}-\ldots-c_{\Upsilon}u_{\Upsilon}^{-}\in H_{\beta}(V).
	\end{equation}
	{\rm 2.} The coefficients in \eqref{eq:3.3.7} are computed by the formula 
	\begin{align}\label{eq:rad_coeff}
		c_{j}=i\big((f,Y_{j}^{-})_{H(\mathring{V})}+(g,-B^N Y_{j}^{-})_{H(\partial V)}\big),
	\end{align}
	where $Y_{j}^{-}$ are the functions \eqref{eq:3.3.8}.\\
	{\rm 3.} The solution $u$ is determined up to an arbitrary term in $\ker\mathcal{L}_{\beta}(\mu)$, and the inequality
	\begin{equation}\label{eq:3.3.9}
		\|v;H_{\beta}(V)\|+|c_{1}|+\ldots+|c_{\Upsilon}|\leqslant \mathrm{const}(\|f;H_{\beta}(\mathring{V})\|+\|g;H_{\beta}(\partial V)\|+\|v; H_{\beta}(W)\|)
	\end{equation}
	holds, where $W$ is a finite subset of $V$. The solution $u_{\perp}$ satisfying the additional orthogonality conditions $(u_{\perp},z_{j})_{H(V)}=0$, $j=1,\ldots,\Theta$, is unique, and for it the inequality \eqref{eq:3.3.9} holds without the last term in the right-hand side.
\end{thm}

\begin{proof}
	By Proposition \ref{pr:coker2}, the conditions \eqref{eq:sol_cond} ensure the existence of a solution $u\in H_{-\beta}(V)$ to the system \eqref{eq:3.1.1} with the right-hand side under consideration. Since the functions $z_{1},\ldots,z_{\Theta}$, $Y_{1}^{+},\ldots,Y_{\Upsilon}^{+}$ form a basis in the space $\ker\mathcal{L}_{-\beta}(\mu)$, the general solution can be written as
	\begin{equation}\label{eq:3.3.10}
		u=u_{0}+\sum\limits_{j=1}^{\Upsilon}a_{j}Y_{j}^{+}+\sum\limits_{k=1}^{\Theta}b_{k}z_{k},
	\end{equation}
	where $u_{0}\in H_{-\beta}(V)$ is a particular solution, and the constants $a_{j}$ and $b_{k}$ are arbitrary. The function $u_{0}$ solves the system \eqref{eq:3.1.1} with the right-hand side $\{f,g\}\in \cH_{\beta}(\mathring{V}, \partial V)$ and therefore meets the conditions of Theorem \ref{th:3.2.1} in all cylindrical outlets $\Pi^{+}_r$, $r=1,\ldots,\mathcal{T}$. Hence
	\begin{equation}\label{eq:3.3.11}
		u_{0}=\sum\limits_{j=1}^{\Upsilon}\tilde{c}_{j}^{+}u_{j}^{+}+\sum\limits_{j=1}^{\Upsilon}\tilde{c}_{j}^{-}u_{j}^{-}+v_{0},
	\end{equation}
	where $v_{0}\in H_{\beta}(V)$, $\tilde{c}_{j}^{+},\;\tilde{c}_{j}^{-}$ are some constants, $j=1,\ldots,\Upsilon$. According to formula \eqref{eq:3.3.6},  the functions $Y_{j}^{+}$, $j=1,\ldots,\Upsilon$, can be represented as
	\begin{equation}\label{eq:3.3.12}
		Y_{j}^{+}=u_{j}^{+}+\sum\limits_{l=1}^{\Upsilon}u_{l}^{-}S_{lj}+v_{j}
	\end{equation}
with $v_{j}\in H_{\beta}(V)$. Then, substituting \eqref{eq:3.3.11} and \eqref{eq:3.3.12} into  \eqref{eq:3.1.10} and choosing $a_{j}=-\tilde{c}_{j}^{+}$, we get
	$$
	u-\sum\limits_{l=1}^{\Upsilon}u_{l}^{-}\Big(\tilde{c}_{l}^{-}-\sum\limits_{j=1}^{\Upsilon}S_{lj}\tilde{c}_{j}^{+}\Big)=v_{0}-\sum\limits_{j=1}^{\Upsilon}\tilde{c}_{j}^{+}v_{j}+\sum\limits_{k=1}^{\Theta}b_{k}z_{k}\in H_{\beta}(V).
	$$
	Thus the solution $u$ with arbitrary coefficients $b_{k}$ satisfies the radiation conditions \eqref{eq:3.3.7}. Item 1 is proved.
	
	We proceed to the proof of item 2. 
	Let us compute $iq(u,Y_{j}^{-})$ in two ways. According to \eqref{eq:3.3.4}, the functions $u$ and $Y_{j}^{-}$ can be replaced by their asymptotics \eqref{eq:3.3.7} and \eqref{eq:3.3.13}. Then, by \eqref{eq:3.3.5}, we have
	$$
	\begin{gathered}
		iq(u,Y_{j}^{-})=iq\left(\sum\limits_{k=1}^{\Upsilon}c_{k}u_{k}^{-},\,u_{j}^{-}+\sum\limits_{l=1}^{\Upsilon}u_{l}^{+}(S^{-1})_{lj}\right)=c_{j}.
	\end{gathered}
	$$
	On the other hand, using the definition \eqref{eq:3.3.2} of the form $q$, we obtain
	\begin{align}
		q(u,Y_{j}^{-})=&((A-\mu)u,Y_{j}^{-})_{H(\mathring{V})}+(u,-B^N  Y_{j}^{-})_{H(\partial V)}-\nonumber\\
		&-(u,(A-\mu)Y_{j}^{-})_{H(\mathring{V})}-(-B^N u, Y_{j}^{-})_{H(\partial V)}=\nonumber\\
		=&(f,Y_{j}^{-})_{H(\mathring{V})}+(g,-B^N  Y_{j}^{-})_{H(\partial V)}.
	\end{align}
	In the second equality we used that $u$ is a solution of the system \eqref{eq:3.1.1}, and $Y_{j}^{-}$ is a solution of the corresponding homogeneous system.
	
	Finally, we proceed to item 3. From \eqref{eq:rad_coeff} it follows that
	\begin{equation}\label{eq:3.3.14}
		|c_{j}|\leqslant \|f\|_{H_{\beta}(\mathring{V})}\|Y_{j}^{-}\|_{H_{-\beta}(\mathring{V})}+\|g\|_{H_{\beta}(\partial V)}\|B^N Y_{j}^{-}\|_{H_{-\beta}(\partial V)}\leqslant\mathrm{const}\big(\|f\|_{H_{\beta}(\mathring{V})}+\|g\|_{H_{\beta}(\partial V)}\big).
	\end{equation}
	By the a priori estimate \eqref{eq:3.1.8},
	\begin{equation}\label{eq:3.3.15}
		\|v;H_{\beta}(V)\|\leqslant {\rm const}(\|(A-\mu)v;H_{\beta}(\mathring{V})\|+\|v;H_{\beta}(\partial V)\|+\|v;H_{\beta}(W)\|),
	\end{equation}
	where $W$ is a finite subset of $V$. Next, by \eqref{eq:3.3.7}, we have
	\begin{align}\label{eq:3.3.16}
		\|(A-\mu)v;H_{\beta}(\mathring{V})\|&\leqslant \|f;H_{\beta}(\mathring{V})\|+\sum\limits_{j=1}^{\Upsilon}|c_{j}|\cdot\|(A-\mu)u_j^-;H_{\beta}(\mathring{V})\|,\nonumber\\
		\|v;H_{\beta}(\partial V)\|&\leqslant \|g;H_{\beta}(\partial V)\|+\sum\limits_{j=1}^{\Upsilon}|c_{j}|\cdot\|u_j^-;H_{\beta}(\partial V)\|.
	\end{align}
	By construction, $u_j^-=0$ on $\partial V$, so $\|u_j^-;H_{\beta}(\partial V)\|=0$. We show that the norm $\|(A-\mu)u_j^-;H_{\beta}(\mathring{V})\|$ is finite. We have on $\mathring{V}$
	$$
	(A-\mu)u_j^-=(A^{(0)}-\mu)u_{j}^{-}+\Delta A u_{j}^{-},
	$$
	where $A^{(0)}$ is the operator introduced after formula \eqref{eq:3.1.4} and $\Delta A=A^{(0)}-A$.
	By construction, $(A^{(0)}-\mu)u_{j}^{-}=0$ has a finite support, and $\Delta A u_{j}^{-}\in H_{\beta}(\mathring{V})$ by the stabilization conditions \eqref{eq:3.2.1}. Estimate \eqref{eq:3.3.9} now follows from \eqref{eq:3.3.14}-\eqref{eq:3.3.16}.
	
	The solution $u_{\perp}$ satisfying the additional orthogonality conditions from item 3 is unique. When deriving the inequality \eqref{eq:3.3.15} for $v$ corresponding to $u_{\perp}$, the a priori estimate \eqref{eq:3.1.8} is replaced by \eqref{eq:3.1.21}, and therefore the last term in \eqref{eq:3.3.9} can be omitted.
\end{proof}


\begin{thebibliography}{99}
 
 
 \bibitem{EgorKopTes}
 I.	Egorova, M. Holzleitner, and G. Teschl, Properties of the scattering matrix and dispersion estimates for Jacobi operators, \emph{J. of Math. Anal. and Appl.},  Vol. 434 (2016), No. 1, pp. 956-966.
 
 
 \bibitem{Su13} 
 T. Sunada, \textit{Topological Crystallography: With a View towards Discrete Geometric Analysis}, Surveys and Tutorials in the Applied Mathematical Sciences, Vol. 6. Springer. Tokyo, 2013.
 
 \bibitem{HS} 
 Y. Higuchi and T. Shirai, The spectrum of magnetic Schr\"{o}dinger operators on a graph with periodic structure, \textit{Journal of Functional Analysis}, Vol. 169 (1999), 456–480.
 
 
  \bibitem{KorSab1}
 E. Korotyaev and N. Saburova, Spectral estimates for Schr\"odinger operators on periodic discrete graphs,  \textit{St. Petersburg Mathematical Journal}, Vol. 30 (2019), No. 4, pp 667–698.

 
   \bibitem{SloKor}
E. Korotyaev and V.Sloushch, Asymptotics and estimates for the discrete spectrum of the Schr\"odinger operator on a discrete periodic graph, \textit{St. Petersburg Math. J.}, 32:1 (2021), pp 9–29.
 

 \bibitem{AndIsoMar}
 	K. Ando, H. Isozaki, and H. Morioka, Spectral properties of Schr\"{o}dinger operators on perturbed lattices, \textit{Annales Henri Poincar\'{e}}, Vol. 17 (2016), No. 8, pp. 2103--2171.
 
  \bibitem{ParaRich}
D. Parra and S. Richard, Spectral and scattering theory for Schr\"odinger operators on perturbed topological crystals, \textit{Reviews in Mathematical Physics}, Vol. 30 (2018), No. 4.
 
 
  \bibitem{Para} 
  D. Parra, Spectral and scattering theory for Gauss–Bonnet operators on perturbed topological crystals, \textit{Journal of Mathematical Analysis and Applications}, Vol. 452 (2017), No. 2, pp 792-813.
 
 
 \bibitem{BerKuch}
G.~Berkolaiko and P.~Kuchment, \emph{Introduction to quantum graphs}, Mathematical Surveys and Monographs, Vol.~186, American Mathematical Society, Providence, RI, 2013.

 
  \bibitem{NazPlam}
S. Nazarov and B. Plamenevsky, \emph{Elliptic Problems in Domains with Piecewise Smooth Boundaries}, de Gruyter Expositions in Mathematics, Vol.~13, Walter de Gruyter \& Co., Berlin, 1994; Russian orig.: Nauka, Moscow, 1991. 

 
  \bibitem{Kato}
T. Kato, \textit{Perturbation Theory for Linear Operators}, Classics in Mathematics, Springer-Verlag, Berlin, 1995.

 
  \bibitem{KozMaz}
  Kozlov V., Maz'ya V., \textit{Differential equations with operator coefficients: with applications to boundary value problems for partial differential equations}, Springer Science \& Business Media, 2013.
 
 
  \bibitem{Her}
L.~H{\"o}rmander, {\it An Introduction to Complex Analysis in Several Variables}, D. Van Nostrand Company, Inc., Princeton, N.J., 1966.
 
 
\end{thebibliography}
\end{document}